\documentclass[twocolumn]{svjour3}

\smartqed  % flush right qed marks, e.g. at the end of the proof

\usepackage{graphicx}
\usepackage[utf8]{inputenc}
\usepackage[T1]{fontenc}
\usepackage{hyperref}
\usepackage{url}
\usepackage{xcolor}
\usepackage{booktabs}
\usepackage{amsfonts}
\usepackage{amsmath}
\usepackage{amssymb}
\usepackage{enumitem}
\usepackage{nicefrac}
\usepackage{microtype}
\usepackage{moresize}
\usepackage{balance}
\usepackage{subcaption}
\usepackage{caption}
\usepackage{lmodern}
\usepackage[noend, linesnumbered, ruled]{algorithm2e}
\SetKwIF{If}{ElseIf}{Else}{if}{}{else if}{else}{end if}%
\SetKwFor{While}{while}{}{end while}%
\SetKwFor{For}{for}{}{end for}
\SetKwRepeat{Do}{do}{while}

\usepackage{amsmath,amsfonts,bm}

\newcommand{\Endure}{{\sc Endure}}
\newcommand{\cost}{{{\ensuremath{\text{C}}}}}
\newcommand{\costvec}{{{\ensuremath{\bf{c}}}}}

\newcommand{\configuration}{{{\ensuremath{\Phi}}}}

\newcommand{\sizeratio}{{{\ensuremath{T}}}}
\newcommand{\mbuf}{{{\ensuremath{m_{\mathrm{buf}}}}}}
\newcommand{\mfilt}{{{\ensuremath{m_{\mathrm{filt}}}}}}
\newcommand{\mtot}{{{\ensuremath{m}}}}
\newcommand{\policy}{{{\ensuremath{\pi}}}}

\newcommand{\asym}{{{\ensuremath{f_{a}}}}}
\newcommand{\querySel}{{{\ensuremath{S_\mathrm{RQ}}}}}

\newcommand{\robustrho}{{{\ensuremath{\bf{\rho}}}}}

\newcommand{\workload}{{{\ensuremath{\bf{w}}}}}

\newcommand{\columnvec}{{{\ensuremath{\bf{e}}}}}

\newcommand{\obsworkload}{{{\ensuremath{\hat{\bf{w}}}}}}

\newcommand{\nonemptylookup}{{{\ensuremath{z_1}}}}
\newcommand{\emptylookup}{{{\ensuremath{z_0}}}}
\newcommand{\range}{{{\ensuremath{q}}}}
\newcommand{\update}{{{\ensuremath{w}}}}
\newcommand{\benchmark}{{{\ensuremath{\mathcal{B}}}}}

\DeclareMathOperator*{\argmax}{arg\,max}
\DeclareMathOperator*{\argmin}{arg\,min}

\newcommand{\nominal}{{\sc Nominal Tuning}}
\newcommand{\robustw}{{\sc Robust Tuning}}

\newcommand{\etal}{\emph{et al.}}

\newcommand\Paragraph[1]{\vspace{0.02in}  \noindent \textbf{#1.}}

\newcommand{\squishlist}{\begin{list}{$\bullet$}
  { \setlength{\itemsep}{0pt}
     \setlength{\parsep}{3pt}
     \setlength{\topsep}{3pt}
     \setlength{\partopsep}{0pt}
     \setlength{\leftmargin}{1.5em}
     \setlength{\labelwidth}{1em}
     \setlength{\labelsep}{0.5em} } }
\newcommand{\squishend}{
  \end{list}  }

\newcommand{\kmodel}{K-LSM}
\newcommand{\dostmodel}{Dostoevsky}
\newcommand{\fluidmodel}{Fluid LSM}
\newcommand{\llmodel}{Lazy Leveling}

\newcommand{\onelmodel}{1-Leveling}

\begin{document}

\title{Towards Flexibility and Robustness of LSM Trees
}
% \subtitle{Do you have a subtitle?\\ If so, write it here}
% \titlerunning{Short form of title}        % if too long for running head

\author{Andy Huynh           \and
        Harshal A. Chaudhari \and
        Evimaria Terzi       \and
        Manos Athanassoulis
}
%\authorrunning{Short form of author list} % if too long for running head
\institute{Andy Huynh \at
                Boston University \\
                \email{ndhuynh@bu.edu}
           \and
           Harshal A. Chaudhari \at
               Boston University \\
              \email{harshal@bu.edu}
           \and
           Evimaria Terzi \at
               Boston University \\
               \email{evimaria@bu.edu}
           \and
           Manos Athanassoulis \at
               Boston University \\
               \email{mathan@bu.edu}
}

\date{Received: date / Accepted: date}
% The correct dates will be entered by the editor
 
\maketitle

\begin{abstract}
Log-Structured Merge trees (LSM trees) are increasingly used as part of the storage engine behind several data systems, and are frequently deployed in the cloud.
As the number of applications relying on LSM-based storage backends increases, the problem of performance tuning of LSM trees receives increasing attention.
We consider both \emph{nominal} tunings -- where workload and execution environment are accurately known \emph{a priori} -- and \emph{robust} tunings -- which consider \emph{uncertainty} in the workload knowledge.
This type of workload uncertainty is common in modern applications, notably in shared infrastructure environments like the public cloud.

To address this problem, we introduce {\Endure}, a new paradigm for tuning LSM trees in the presence of workload uncertainty.
Specifically, we focus on the impact of the choice of compaction policy, size ratio, and memory allocation on the overall performance.
{\Endure} considers a robust formulation of the throughput maximization problem and recommends a tuning that offers near-optimal throughput when the executed workload is not the same, instead in a \emph{neighborhood} of the expected workload.
Additionally, we explore the robustness of flexible LSM designs by proposing a new unified design called {\kmodel} that encompasses existing designs.
We deploy our robust tuning system, {\Endure}, on a state-of-the-art key-value store, RocksDB, and demonstrate throughput improvements of up to 5{$\times$} in the presence of uncertainty.
Our results indicate that the tunings obtained by {\Endure} are more robust than tunings obtained under our expanded LSM design space.
This indicates that robustness may not be inherent to a design, instead, it is an outcome of a tuning process that explicitly accounts for uncertainty.
\end{abstract}

% =============================================================================
\section{Introduction}
\label{sec:introduction}
% =============================================================================

\Paragraph{Ubiquitous LSM-based Key-Value Stores}
Log-Structured Merge trees (LSM trees) are commonly deployed as the backend storage engine of modern key-value stores~\cite{ONeil1996}.
The high ingestion rates and fast reads provided by LSM trees have led to their wide adoption by systems like RocksDB~\cite{FacebookRocksDB} at Meta, LevelDB~\cite{GoogleLevelDB} and BigTable~\cite{Chang2006} at Google, HBase~\cite{HBase2013}, Cassandra~\cite{ApacheCassandra} at Apache, WiredTiger~\cite{WiredTiger-a} at MongoDB, X-Engine~\cite{Huang2019} at Alibaba, and DynamoDB~\cite{DeCandia2007} at Amazon.

LSM trees store incoming data within a memory buffer, which is subsequently flushed to storage when full, and merged with earlier buffers to form a collection of sorted runs with exponentially increasing sizes~\cite{Luo2020b}.
Frequent merging of sorted runs leads to a higher merging cost, but facilitates faster lookups (\emph{leveling}).
On the flip side, lazy merging policies (\emph{tiering}) trade lookup performance for lower merging costs~\cite{Sarkar2021c}.

\Paragraph{Tuning LSM trees}
As the number of applications relying on LSM-based storage backends increases, the problem of performance tuning LSM trees has garnered a lot of attention.
A common assumption of these methods is that when creating an instance-optimized system~\cite{Kraska2021}, one has \emph{complete knowledge of the expected workload and the execution environment}.
Given such knowledge, prior work focuses on optimizing LSM tree parameters such as memory allocation for Bloom filters across different levels, memory distribution between the buffers and the Bloom filters, and the choice of merging policies (i.e., \emph{leveling} or \emph{tiering})~\cite{Dayan2017}.
Different optimization objectives have led to hybrid merging policies with more fine-grained tunings~\cite{Dayan2018,Dayan2019,Idreos2019}; optimized memory allocation strategies~\cite{Bortnikov2018,Kim2020,Luo2020a}, Bloom filter variations~\cite{Luo2020,Zhang2018a}, new compaction strategies~\cite{Alkowaileet2020,Luo2019a,Sarkar2020,Sarkar2021c,Zhang2020}, and exploitation of data characteristics~\cite{Absalyamov2018,Ren2017,Yang2020}.

Even when accurate information about the workload and underlying hardware are available, tuning data systems is a notoriously difficult research problem~\cite{Chaudhuri2004a,Chaudhuri2005,Shasha2002}.
Additionally, the explosive growth in the usage of the cloud infrastructure for data management~\cite{Hayes2008,GrandViewResearch2019} has exacerbated this problem due to the increase in uncertainty and variability in workloads~\cite{Chohan2010,Galante2012,Herbst2013,Holze2010,Ozcan2017,Pezzini2014,Schnaitter2006,Schnaitter2007,Schnaitter2012,Wolski2017}.

To address this challenge, we introduce \Endure\footnote{An earlier version of this work appeared in VLDB 2022~\cite{Huynh2022}.}, a general framework for providing robust tunings under uncertain input workloads.
{\Endure} introduces a tuning-under-uncertainty paradigm by formulating the classic tuning problem as a robust optimization problem.
Our experiments demonstrate the benefits of robust tunings compared to existing baselines for tuning LSM trees.

\Paragraph{Expanding the LSM Design Space}
In this paper, we build on our prior work~\cite{Huynh2022} and expand it along multiple dimensions.
We take a more critical look at the performance of LSM tunings for flexible LSM designs both with and without workload uncertainty.
After careful consideration of existing LSM designs and tuning approaches---e.g., Monkey~\cite{Dayan2017} and Dostoevsky~\cite{Dayan2018}---we propose a general and more unified LSM design, termed {\kmodel}.
Our design allows each level to have a variable number of potentially overlapping files.
Therefore, we can describe both standard compaction policies (i.e., tiering~\cite{Jagadish1997} and leveling~\cite{ONeil1996}), and existing hybrid compaction policies.
We demonstrate that {\kmodel} can reduce to data layouts such as {\llmodel}~\cite{Dayan2019,Dayan2018}, {\dostmodel}~\cite{Dayan2018}, and 1-Leveling~\cite{Sarkar2021c}, thus making it a unified LSM design.
Furthermore, we accompany the {\kmodel} design with a cost model, which in turn can capture the costs of all aforementioned approaches.

\Paragraph{Performance of LSM Tunings}
Next, we check the feasibility of tuning an LSM tree under the assumption of an accurately known workload (no uncertainty) using the {\kmodel} cost model.
We show that this can be done using off-the-shelf numerical solvers.
Our experiments indicate that tunings obtained using flexible designs provide better system performance when compared to those obtained from state-of-the-art LSM designs.
To the best of our knowledge, we are the first to propose a unified LSM cost model.

\Paragraph{Robustness of LSM Trees}
In the second part of the work, we present results with {\Endure}~\cite{Huynh2022}, our system for \emph{robust} LSM tree tuning---i.e., LSM tree tuning in the presence of workload uncertainty.
Here, we depart from the classical view of database tuning, which assumes accurate knowledge about the expected workload.
Towards this, {\Endure} introduces a new \emph{robust tuning paradigm} that incorporates expected uncertainty into optimization and applies it to LSM trees.

We formulate the {\robustw} problem that seeks an LSM tree configuration that maximizes the worst-case throughput over all the workloads in the \emph{neighborhood} of an expected workload.
We use the notion of KL-divergence between probability distributions to define the neighborhood size, implicitly assuming that the uncertain workloads would be contained in the neighborhood.
As the KL-divergence boundary condition approaches zero, our problem becomes equivalent to the classical optimization problem (henceforth referred to as the {\nominal} problem).
More specifically, our approach uses as input the expected size of the uncertainty neighborhood, which dictates the qualitative characteristics of the solution.
Intuitively, the larger the size of the uncertainty neighborhood, the larger the workload discrepancy a robust tuning can accommodate.
Leveraging work on robust optimization from the Operations Research and Machine Learning communities~\cite{Ben-Tal2013,Ben-Tal1998-a,Bertsimas2010-a}, we efficiently solve the {\robustw} problem and find the robust tuning for LSM tree-based storage systems.
A similar problem of using workload uncertainty while determining the physical design of column-stores has been explored in prior work~\cite{Mozafari2015}.
However, this methodology is not well-suited for the LSM tuning problem.
We provide additional details regarding this in Section~\ref{sec:related_work}.

\Paragraph{Flexibility in Design and Robustness}
Finally, we experimentally investigate whether the nominal tunings obtained by various LSM designs are inherently robust.
That is, we investigate whether the lack of robustness in the state-of-the-art nominal tunings is a consequence of the designs not being expressive enough, or a result of the tuning process's lack of consideration for uncertainty.
Our findings indicate that the nominal tunings obtained via {\kmodel} provide a benefit over traditional LSM designs in scenarios where the workload does not deviate from the expected.
However, this benefit does not appear in the contrasting scenario where the workload does deviate from the expected.
Rather, tunings obtained from {\Endure} exhibit higher throughput with simpler LSM designs than nominal tunings with flexible LSM designs.
Hence, we conclude flexibility does not inherently provide robustness.

\Paragraph{Contributions}
To the best of our knowledge, our work is the first that presents a unified LSM design with an associated cost model that is a generalization of all the existing state-of-the-art approaches.
Moreover, we present the first systematic approach to selecting a robust tuning for instance-optimized LSM tree-based key-value stores under workload uncertainty, utilizing robust optimization techniques from machine learning.
Finally, we are the first to explore whether the robustness of an LSM tree can be an inherent design property or a result of explicitly tuning for uncertainty.

Our technical and practical contributions can be summarized as follows:
\begin{itemize}[leftmargin=1.3em, itemsep=0.22em]
    \item We introduce {\kmodel}, a new unified LSM design that describes both classic designs and recently proposed state-of-the-art hybrid designs (\S\ref{sec:lsm-cost-model}).
        We present its implications on classical LSM tuning (\S\ref{sec:tuning-lsm-trees}).
    \item We incorporate workload uncertainty into LSM tunings and provide algorithms to compute robust tunings efficiently.
        {\Endure} can be tuned for varying degrees of workload uncertainty, and is practical enough to be adopted by the current state-of-the-art LSM storage engines (\S\ref{sec:robust-tuning}).
    \item We develop an uncertainty benchmark that can evaluate the robustness of the current state-of-the-art LSM-based systems (\S\ref{sec:uncertainty-benchmark}).
    \item In our model-based analysis, we show that robust tunings obtained from {\Endure} provide up to 5$\times$ higher throughput when faced with uncertain workloads (\S\ref{sec:model-evaluation}).
    \item We deploy and test {\Endure} in RocksDB, a state-of-the-art LSM storage engine, to demonstrate the feasibility of using robust tunings on commercial systems.
        We show that {\Endure} achieves up to $2.4\times$ throughput speedups, and these gains are independent of the database size (\S\ref{sec:system-evaluation}).
    \item By evaluating the robustness of the {\kmodel} design, we demonstrate that \emph{robustness is not inherent to a design,} rather is an outcome of a tuning process that accounts for uncertainty (\S\ref{sec:robustness-flexible-models}).
    \item To encourage reproducible research, we make all our code publicly available\footnote{\url{https: //github.com/BU-DiSC/endure}}.
\end{itemize}

% =============================================================================
\section{Background on LSM Trees}
\label{sec:background}
% =============================================================================
\begin{figure}[t]
   \centering
   \includegraphics[width=\columnwidth]{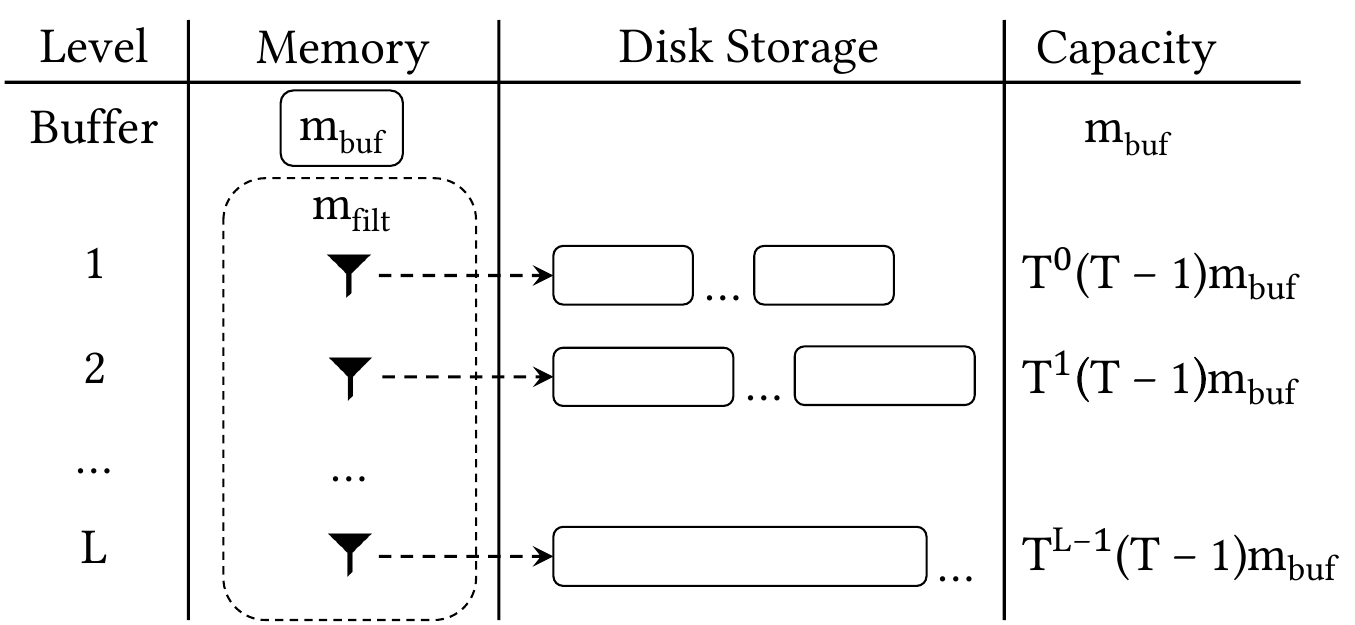}
   \caption{Overview of the structure of an LSM tree}
   \label{fig:lsm-overview}
\end{figure}

\Paragraph{Basics}
LSM trees use the \textit{out-of-place} paradigm to store key-value pairs~\cite{Luo2020b}.
Inserts, updates, and deletes are placed in a memory buffer.
Once full, its contents are sorted based on the key, forming an \textit{immutable sorted run}, then flushed to secondary storage.
Sorted runs are subsequently organized into logical levels.

Thus, for an LSM tree with $L$ disk-resident levels, we label the memory buffer as Level $0$ and the remaining levels in storage from $1$ to $L$.
The disk-resident levels have exponentially increasing sizes dictated by a tunable size ratio $\sizeratio$.
Figure~\ref{fig:lsm-overview} shows an overview of an LSM tree.

We denote the number of bits of main memory allocated to the buffer $\mbuf$, which holds several entries with a fixed entry size $E$.
For example, in RocksDB, the default buffer size is $\mbuf=64$MB, and depending on the application, the entry size typically varies between 64B and 1KB.
The buffer at Level 0 is mutable and can be updated in place, while runs starting at Level 1 and beyond are \emph{immutable}.
Each Level $i$ has a capacity threshold of $(\sizeratio - 1) \cdot\sizeratio^{i-1} \cdot \frac{\mbuf}{E}$ entries, thus, the level capacities are exponentially increasing by a factor of $\sizeratio$.
The total number of levels $L$ for a given {\sizeratio} is
\begin{equation}
L(\sizeratio) =
    \Bigg\lceil
        \log_\sizeratio \left( \frac{N \cdot E}{\mbuf} + 1 \right)
    \Bigg\rceil ,
\label{eq:levels}
\end{equation}
where $N$ is the total number of entries~\cite{Dayan2017,Luo2020,Sarkar2020}.

\Paragraph{Compaction Policies: Leveling and Tiering}
Classically, LSM trees support two compaction policies: leveling and tiering~\cite{Luo2020b,Sarkar2022b}.
In leveling, each level contains at most one run, and every time a run in Level $i - 1$ ($i \geq 1$) is flushed to Level $i$, it greedily sort-merges (compacts) with the run from Level $i$, assuming it exists.
With tiering, every level must accumulate $\sizeratio$ runs before a compaction is triggered.
During a compaction, entries with a matching key are consolidated, and only the most recent valid entry is retained~\cite{Dong2017,ONeil1996,Sarkar2021c}.

\Paragraph{Flexible Compaction Policies}
The different LSM tree compaction policies form a continuum between a read-optimized and write-optimized data layout, where leveling and tiering policies are the two extremes~\cite{Sarkar2022b}.
Hybrid compaction policies allow a smooth transition of the tree shape to strike a balance between the read and write throughput~\cite{Dayan2018,Dayan2019}.
{\llmodel} assigns the upper levels of the LSM tree to a tiering policy and the last level to a leveling policy to improve the worst-case cost for writes while maintaining near-optimal read performance.
This is motivated by the fact that the last level statistically contains most of the LSM tree's data.

This notion of assigning different compaction policies per level is further expanded by the Dostoevsky design and the Fluid LSM tree~\cite{Dayan2018}.
Rather than assigning each level a different compaction policy, the {\fluidmodel} tree uses two limits for the number of runs per level, one for the last level of the LSM tree, and a different one for all the upper levels.
This allows the {\fluidmodel} design to express fine-grained hybrid compaction policies between leveling and tiering.

In this work, we further expand this approach by proposing {\kmodel}, a more expressive LSM compaction model that unifies all prior approaches by allowing each level to parameterize its capacity in terms of the number of files it can hold.
In Section~\ref{sec:lsm-cost-model}, we discuss {\kmodel} in detail and demonstrate that it can explore a wider design space.
We further discuss its implications on the robustness of its tunings.

\Paragraph{LSM tree Operations}
An LSM tree supports three basic operations: (a) writes of new key-value pairs, (b) point queries, and (c) range queries.

\emph{Writes:}
All write operations are handled by a buffer append.
Once the buffer is full, a compaction is triggered.
Any write may include either a new key-value pair, an existing key that \emph{updates} its value, or a special entry called a tombstone that \emph{deletes} an existing key.

\emph{Point Queries:}
A point query searches for the value of a specific key.
It begins by looking at the memory buffer and then traverses the tree from the smallest to the largest level.
At each level, the lookup moves from the most recent sorted run to the oldest sorted run, terminating when it finds the first matching entry.
Note that a point query might return either an \emph{empty} or a \emph{non-empty} result.
We differentiate the two as it has been shown workloads with empty point queries can be further optimized~\cite{Dayan2017}.

\emph{Range Queries:}
A range query lookup returns the most recent version of all keys within the desired range by potentially scanning every run at every level.

\Paragraph{Optimizing Lookups}
LSM tree lookups are optimized using \emph{filters} and indexes (also termed \emph{fence pointers})~\cite{Sarkar2023}.
In the worst case, a lookup needs to probe every run, however, LSM engines use one filter per run~\cite{Dayan2017,FacebookRocksDB} to reduce this cost.
While the filters are part of each run, they are aggressively cached in memory.
One of the most common filter designs used in LSM trees is the Bloom filter~\cite{Bloom1970}.
A Bloom filter is a probabilistic membership test data structure that responds with a false positive rate $f$, which is a function of the ratio between the number of memory bits allocated {\mfilt} and the number of elements indexed.
By probing the Bloom filter of a particular level, an LSM tree can skip accessing that run altogether when it does not contain the indexed key.
In practice, for efficient storage, Bloom filters are maintained at the granularity of files~\cite{Dong2017}.
Fence pointers hold the smallest key for each disk page of all sorted runs into main memory~\cite{Dayan2017} to quickly identify which page(s) to read for a lookup.
In this work, we assume that fence pointers are required and consume a fixed amount of memory in the system.
Therefore, any operation that requires a single I/O will only require one logical page lookup by the operating system by following the corresponding fence pointer.
We further assume that a single I/O operation corresponds to exactly one logical page access.

\Paragraph{Tuning LSM Trees}
An LSM tree is a highly tunable data structure where the size ratio, compaction policy, exact shape of the tree, and memory allocation can all be tuned.
Classical LSM tuning strategies start with an offline analysis and assume the workload information and the execution environment are accurately known a priori to deployment.
In comparison, online tuning strategies change LSM tuning knobs in response to workloads, however, the design parameters that mainly drive the performance must be dictated before deployment~\cite{Kim2017a,Luo2020b}.
While LSM trees are also deployed as collections that can be co-tuned~\cite{Luo2020c}, here we focus on deploying and tuning single instances of LSM trees.
Under that assumption, LSM tree tuning considers the optimal allocation of available main memory between Bloom filters and buffering~\cite{Kim2020,Luo2020a}, the optimal choice of size ratio, and the data layout strategy~\cite{Dayan2017,Dayan2018a,Dayan2018}.
Such design decisions are common across industry-standard LSM-based engines such as Apache Cassandra~\cite{ApacheCassandra}, AsterixDB~\cite{Alsubaiee2014}, RocksDB~\cite{FacebookRocksDB}, and InfluxDB~\cite{Influxdata2021-a}.
Lastly, recent work has introduced new hybrid merging strategies~\cite{Dayan2019,Sarkar2021c}, and optimizations for faster data ingestion~\cite{Luo2019b} and performance stability~\cite{Luo2019a}.

\section{Preliminaries}\label{sec:notation}
As we discussed above, LSM trees have two types of parameters: the \emph{design parameters} that are changed primarily for performance, and the \emph{system parameters} that are a part of the system the LSM tree is deployed on, and therefore untunable.

\Paragraph{Design Parameters}
The design parameters we consider in this paper are the size ratio ({\sizeratio}), the memory allocated to the Bloom filters ({\mfilt}), the memory allocated to the write buffer ({\mbuf}), and the compaction policy ({\policy}).
These are ubiquitous design parameters and have been extensively studied as having the largest impact on performance~\cite{Dayan2017,Luo2020b}.
Therefore, we focus on these parameters to define a problem that is not bound to any specific LSM engine.
Recall that the compaction policy refers to either leveling or tiering in a classical design, or may contain other parameters used to describe hybrid designs as we discuss in Section \ref{subsec:klsm-model}.

\Paragraph{System Parameters}
In production deployments, performance depends on various \emph{system parameters} (e.g., total memory {\mtot}, page size $B$), and other non-tunable data-dependent parameters (e.g., data entry size $E$, amount of data $N$).
We assume these parameters are known a priori and fixed throughout the tuning process.

\Paragraph{LSM Tree Configuration}
We use $\configuration$ to denote the LSM tree tuning configuration which describes the values of the tunable parameters together $\configuration := (\sizeratio, \mfilt, \policy)$.
Note that we only use the memory for Bloom filters $\mfilt$ and not $\mbuf$, because the latter can be derived using the total available memory ($\mbuf=\mtot-\mfilt$).

\Paragraph{Workload}
The choice of the parameters in $\configuration$ depends on the input (expected) workload, i.e., the fraction of empty lookups ({\emptylookup}), non-empty lookups ({\nonemptylookup}), range lookups ({\range}), and write ({\update}) queries within an observation period.
Such a period is defined either over a fixed time interval or over a certain number of queries.
Note that this workload representation is common for analyzing and tuning LSM trees~\cite{Dayan2017,Luo2020b}.
Additionally, complex workloads (i.e., SQL statements) generate access patterns of the storage engine and can be broken down into the same basic operations.
This mapping of complex queries to basic operations is also common for performance tuning of LSM tree-based storage engines~\cite{Cao2020}.
Therefore, a workload can be expressed as a vector $\workload = (\emptylookup, \nonemptylookup, \range, \update)^\intercal \geq 0$ describing the proportions of the different kinds of queries.
Clearly, $\emptylookup+\nonemptylookup+\range+\update = 1$ or alternatively: $\workload^\intercal \columnvec = 1$ where {\columnvec} denotes a column vector of ones of the same dimension as {\workload}.

\begin{table}[t]
    \centering
    \caption{Summary of problem notation}
    \label{tab:params}
    \resizebox{\columnwidth}{!}{
    \begin{tabular}{ccl}
        \toprule
        Type & Term & Definition \\
        \toprule
            Design & $\mfilt$ &  Memory allocated for Bloom filters  \\
                   & $\mbuf$ &  Memory allocated for the write buffer \\
                   & $\sizeratio$ &  Size ratio between consecutive levels   \\
                   & \policy & Compaction policy (\emph{tiering}/\emph{leveling}) \\
        \midrule
            System & $\mtot$ & Total memory (filters+buffer) ($\mtot=\mbuf+\mfilt$)\\
                   & $E$ &  Size of a key-value entry  \\
                   & $B$ &  Number of entries that fit in a page   \\
                   & $N$ & Total number of entries    \\
        \midrule
            Workload & $z_0$ & Percentage of zero-result point lookups\\ %in the workload \\
                     & $z_1$ & Percentage of non-zero-result point lookups\\ %in the workload  \\
                     & $q$ & Percentage of range queries\\ %in the workload  \\
                     & $w$ & Percentage of writes\\ %in the workload  \\\
        \bottomrule
    \end{tabular}
    }
\end{table}

Each type of query (non-empty lookups, empty lookups, range lookups, and writes) has a different cost, denoted as $Z_0(\configuration)$, $Z_1(\configuration)$, $Q(\configuration)$, $W(\configuration)$, as there is a dependency between the cost of each type of query and the design ${\configuration}$.
For ease of notation, we use $\costvec(\configuration) = \left(Z_0(\configuration), Z_1(\configuration), Q(\configuration), W(\configuration)\right)^\intercal$ to denote the vector of the costs of executing different types of queries.
Thus, given a specific configuration ({\configuration}) and a workload ({\workload}), the expected cost can be computed as:
{%
\begin{equation}
\label{eq:thecost}
    \begin{split}
        \cost(\workload, \configuration)
        = \workload^\intercal\costvec(\configuration)
        = &\nonemptylookup \cdot Z_0(\configuration)
            + \emptylookup \cdot Z_1(\configuration) \\
            &+ \range \cdot Q(\configuration)
            + \update \cdot W(\configuration).
    \end{split}
\end{equation}
}%

\noindent We present a summary of all of our notation in Table~\ref{tab:params}.

% =============================================================================
\section{The Cost Model of LSM Trees}
\label{sec:lsm-cost-model}
% =============================================================================

In this section, we provide a detailed cost model that accurately captures the behavior of a wide collection of LSM compaction strategies, including classical leveling and tiering, as well as hybrid approaches.
Following prior work~\cite{Dayan2017,Luo2020}, we focus on the four types of operations described earlier: point queries that return an empty result, point queries that have a match, range queries, and writes.

\subsection{Model Basics}
When modeling the read cost of LSM trees, a key quantity to accurately capture is the amount of superfluous I/Os that take place.
Although Bloom filters are used to minimize extra I/Os, they allow for a small fraction of false positives.
If the filter returns negative, the target key does not exist in the run, and the lookup skips over the assigned fence pointer saving a single random I/O.
If a filter returns positive, then the target key may exist, so the lookup probes the run at a cost of one I/O.
Then, if the run contains the correct key the lookup terminates.
Otherwise, we have a \emph{false positive} and the lookup continues to probe the next run increasing the number of I/Os.
The false positive rate ($\epsilon$) of a standard Bloom filter that is designed to hold information over $n$ entries using a bit-array of size $m$ is given by~\cite{Tarkoma2012}:
\begin{equation*}
\label{eq:bloom}
    \epsilon = e^{-\frac{m}{n} \cdot \ln(2)^2}.
\end{equation*}
Note that the above equation assumes the use of an optimal number of hash functions in the Bloom filter~\cite{enwiki:1025193696}.

Classically, LSM tree-based key-value stores use the same number of bits-per-entry across all Bloom filters.
This means that a lookup probes on average $O\left(e^{-\nicefrac{\mfilt}{N}}\right)$ of the runs, where $\mfilt$ is the overall amount of main memory allocated to the filters.
As $\mfilt$ approaches 0 or infinity, the term $O\left(e^{-\nicefrac{\mfilt}{N}}\right)$ approaches 1 or 0 respectively.
Here, we build on the state-of-the-art Bloom filter allocation strategy proposed in Monkey~\cite{Dayan2017} that uses different false positive rates for each level of the LSM tree to offer optimal memory allocation; for a size ratio {\sizeratio}, the false positive rate corresponding to the Bloom filter at the level $i$ is given by
\begin{equation}
\label{eq:bloom2}
    f_i(\sizeratio) =
        \frac{\sizeratio^{\frac{\sizeratio}{\sizeratio - 1}}}
            {\sizeratio^{L(\sizeratio) + 1 - i}}
        \cdot e^{-\frac{\mfilt}{N}\ln(2)^2}.
\end{equation}
Additionally, false positive rates for all levels satisfy $0 \leq f_i(\sizeratio) \leq 1$.
It should be further noted that Monkey optimizes false positive rates at individual levels to minimize the worst-case average cost of empty point queries.
Non-empty point query costs, being significantly lower than those of empty point queries, are not considered during the optimization process.

\Paragraph{LSM Tree Design \& System Parameters}
In Section~\ref{sec:notation} we introduced the key design and system parameters needed to model LSM tree performance.
In addition to those parameters, there are two more auxiliary and derived parameters we use: the potential storage asymmetry~\cite{Papon2021a} in reads and writes (\asym) and the expected selectivity of range queries (\querySel).

% ============================================================================
\begin{figure}
    \centering
    \begin{subfigure}[b]{\columnwidth}
        \centering
        \includegraphics[width=\columnwidth]{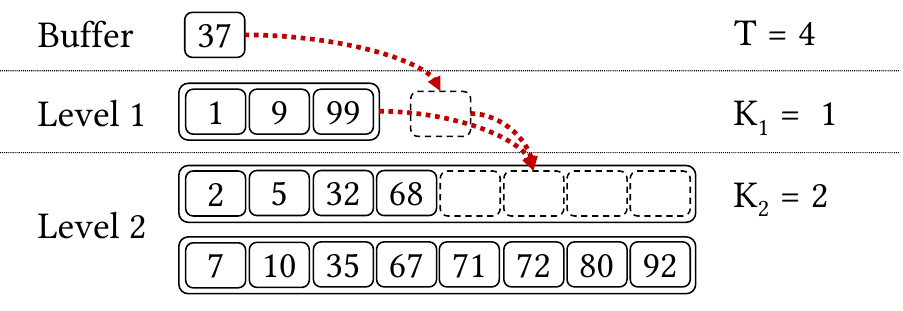}
        \caption{
            Each level has a maximum number of runs before compaction,
            which occurs once the $T^{th}$ run is received.
        }
        \label{fig:klsm_example}
    \end{subfigure}
    \hfill
    \begin{subfigure}[b]{\columnwidth}
        \centering
        \includegraphics[width=\columnwidth]{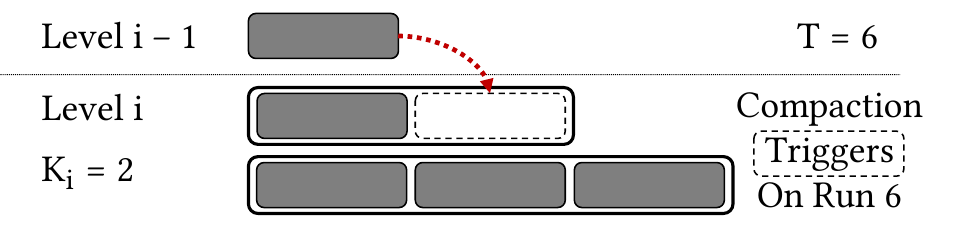}
        \caption{
            Runs do not need to be same size if $\frac{\sizeratio - 1}{K_i}$
            is not integer.
        }
        \label{fig:klsm_uneven_compaction}
    \end{subfigure}
    \caption{
        {\kmodel} provides a flexible way to describe different compaction behaviors.
        In this figure, assume the buffer is the same size as a logical page; then each sorted run is composed of multiple pages.
    }
    \label{fig:klsm_tree}
\end{figure}

\subsection{Extending Classic LSM Compaction Policies}
\label{subsec:klsm-model}
We now introduce a new variable $K_i$ that denotes the maximum number of files for a given level $i$.
It captures a unified design for both classical compaction policies (i.e., tiering and leveling) by introducing a range of new hybrid policies.

\Paragraph{Maximum Files Per Level}
Figure~\ref{fig:klsm_tree} displays the basic structure of an LSM tree with $K_i$ assigned for all levels.
We define $K_i$ as the maximum number of sorted immutable runs before a full-level compaction triggers, essentially the capacity of runs per level.
In a classic tiering compaction policy, a single level of an LSM Tree traditionally has a max of $(\sizeratio - 1)$ runs, each of size $\frac{\mbuf}{E} \cdot \sizeratio^{(i - 1)}$ where $i$ is the assigned level.
A full-level compaction triggers once the level receives $\sizeratio$ runs from the level above, as a result, a level will have at most $(\sizeratio - 1)$ runs.
In our new design, each level still respects the maximum entry capacity for an LSM tree, as each run will have at most $\frac{\mbuf}{E} \cdot \frac{\sizeratio^{i - 1}}{K_i}$ entries.
Figure~\ref{fig:klsm_example} shows an example of a tree right before the compaction occurs.
Once the buffer is flushed, Level 1 will compact all runs within the level and send a sorted run to Level 2, which subsequently sort-merges the received run with the existing data.
Then after four more buffer flushes, Level 2 will have received another run and trigger a compaction, creating a new Level 3.

\Paragraph{Compaction Behavior}
When $K_i = \sizeratio - 1$ for level $i$, the design is equivalent to a tiering policy, while for $K_i = 1$, it is equivalent to a leveling policy.
As incoming sorted runs are compacted, we choose not to split runs, rather, we only merge runs or logically move them from one level to another.
Therefore, for values in between $\sizeratio - 1$ and $1$, we alternate when compacted runs from the level above are merged or simply logically moved.
For example, Figure~\ref{fig:klsm_uneven_compaction} shows a scenario for $K_i = 2$, and $\sizeratio = 6$.
The first three runs from the level above would be compacted to form a single run; the next 2 runs would merge to form a sorted run.
In this instance, each run would be of a different size, one holding the equivalent of three compactions while the other holding 2.
Otherwise, if $K_i$ and $\sizeratio$ are set such that $(K_i - 1) / \sizeratio$ is an integer, the size of each sorted run is equivalent.
The sixth flush from Level $i-1$ triggers a full-level compaction flushing to Level $i+1$.

% ============================================================================

\subsection{A General Cost Model}
\begin{table}[t]
    \centering
    \caption{Additional model notation}
    \label{tab:model_params}
    \resizebox{\columnwidth}{!}{
    \begin{tabular}{cl}
        \toprule
        Term & Definition \\
        \toprule
        $Z_0(\configuration)$ & Empty read cost w.r.t to a specific LSM configuration $\configuration$\\
        $Z_1(\configuration)$ & Non-empty read cost w.r.t to a specific LSM configuration $\configuration$\\
        $Q(\configuration)$ & Range read cost w.r.t to a specific LSM configuration $\configuration$\\
        $W(\configuration)$ & Write cost w.r.t to a specific LSM configuration $\configuration$\\
        $L(T)$ & Number of levels to fill a tree with size ratio $T$\\
        $N_f(T)$ & Number of entries to fill a tree with size ratio $T$\\
        $K_i$ & The maximum number of overlapping files for level $i$   \\ 
        $f_i(T)$ & Bloom filter false positive rate at Level $i$ with size ratio $T$\\
        ${\asym}$ & Read/write Asymmetry ratio for storage device\\
        $f_{seq}$ & Cost of a sequential read w.r.t a random read\\
        ${\querySel}$ & Range query selectivity\\
        \bottomrule
    \end{tabular}
    }
\end{table}
Using the above insights, we model the costs in terms of the expected number of I/O operations required for the fulfillment of the individual queries.
We summarize new notations introduced for the cost model in Table~\ref{tab:model_params}.

\Paragraph{Expected Empty Point Query Cost ($Z_0$)}
A point query that returns an empty result will have visited all sorted runs on every level and issue an I/O for every false positive result amongst the Bloom filters.
Therefore, the expected number of I/Os per level depends on the Bloom filter memory allocation at that level.
Hence, Equation \eqref{eq:read} expresses $Z_0$ in terms of the false positive rates at each level as:
\begin{equation}
\label{eq:read}
    Z_0(\configuration) =
        \sum_{i = 1}^{L(\sizeratio)} K_i \cdot f_i(\sizeratio).
\end{equation}
For each level, there will be at most $K_i$ runs, and each run will have equal false positive rates.

\Paragraph{Expected Non-empty Point Query Cost ($Z$)}
There are two components to the expected non-empty point query cost.
First, we assume that the probability of a point query finding a non-empty result in a level is proportional to the size of the level. 
Thus, the probability of such a query being satisfied on Level $i$ by a unit cost I/O operation is simply $\frac{(\sizeratio-1) \cdot \sizeratio^{i-1}}{N_f(\sizeratio)} \cdot \frac{\mbuf}{E}$, where $N_f(\sizeratio)$ denotes the number of entries in a full tree up to $L(\sizeratio)$ levels:

\begin{equation}
N_f(\sizeratio) =
    \sum_{i = 1}^{L(\sizeratio)}
        (\sizeratio - 1)
        \cdot \sizeratio^{i - 1}
        \cdot \frac{\mbuf}{E}.
\end{equation}

\noindent
Second, we assume that all levels preceding Level $i$ will trigger an I/O operation with a probability equivalent to the false positive rates of the Bloom filters at those levels.
Similarly to empty point queries, the expected cost of such failed I/Os on the preceding levels is $\sum_{j=1}^{i-1}f_j(\sizeratio)$.
Lastly, each level will contain at most $K_i$ sorted runs, we assume that on average the entry is found in the middle run resulting in an additional $\frac{(K_i - 1)}{2} \cdot f_i(\sizeratio)$ extra I/Os.
Thus, we can compute the non-empty point query cost as an expectation over the entry being found at any of the $L(\sizeratio)$ levels of the tree as follows:

\begin{equation}
    \label{eq:read-non-empty}
    \begin{split}
    Z_1(\configuration) &=
        \sum_{i = 1}^{L(\sizeratio)}
            \frac{(\sizeratio - 1) \cdot \sizeratio^{i - 1}}{N_f(\sizeratio)}
            \cdot \frac{\mbuf}{E} \\
        &\cdot \bigg(
            1
            + \sum_{j=1}^{i-1} K_j \cdot f_j(\sizeratio)
            + \frac{K_i - 1}{2} f_i(\sizeratio)
        \bigg).
    \end{split}
\end{equation}
\normalsize

\Paragraph{Range Queries Cost ($Q$)}
A range query will issue at most one disk seek per run per level, or $K_i$ disk seeks.
Each seek is then followed by a sequential scan. The cumulative number of pages scanned over all runs is $\querySel \cdot \frac{N}{B}$, where {\querySel} is the average proportion of all entries included in range lookups.
After finding the first valid page, range queries perform sequential I/Os for subsequent pages rather than a random I/O. 
Therefore, we add a scaling factor $f_{seq}$ that represents the cost of a sequential I/O with respect to one random I/O.
Hence, the overall range lookup cost $Q$ in terms of logical pages reads is as follows: 

\begin{equation}\label{eq:range}
    Q(\configuration) = f_{seq} \cdot \querySel \cdot \frac{N}{B} + \sum_{i=1}^{L(T)} K_i.
\end{equation}

\begin{figure}
    \begin{center}
        \includegraphics[width=\columnwidth]{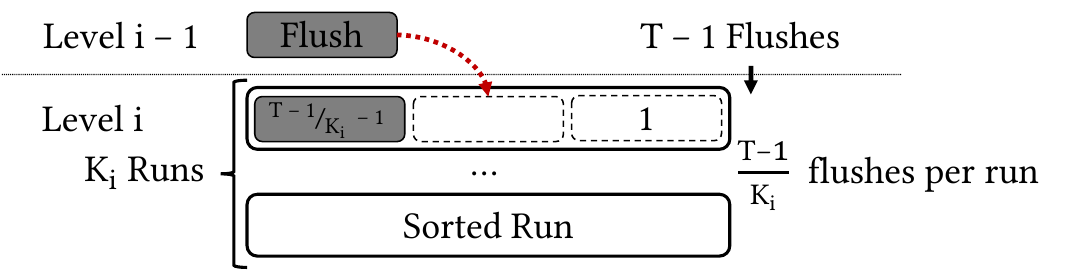}
    \end{center}
    \caption{
        The last flush of a sorted run participates in 1 merge as it eagerly merges into the sorted run.
        The first flush will participate in all subsequent eager merges from new flushes.
    }
    \label{fig:write-cost-derivation}
\end{figure}

\Paragraph{Write Cost ($W$)}
We model worst-case writing cost assuming that the vast majority of incoming entries do not overlap. 
This implies most entries will propagate through all levels of the LSM tree.
Therefore, we calculate the expected number of I/Os by first estimating the average number of merge operations a single write participates in at Level $i$, and summing over all levels.
We start by deriving the total number of merges that occur on Level $i$.
note that Level $i$ will receive at most $\sizeratio - 1$ flushes from Level $i - 1$ before a full level compaction is triggered.
Additionally, a run at Level $i$ needs $\frac{\sizeratio - 1}{K_i}$ flushes from Level $i - 1$ to reach its maximum size; we will refer to this as the flush capacity.
Figure~\ref{fig:write-cost-derivation} shows the number of flushes and flush capacity for Level $i$.

Analyzing a single sorted run at Level $i$, we observe that the last flush will only participate in a single eager compaction as the sorted run will reach its flush capacity at that point.
The second to last flush participates in $2$ merges, the third to last in $3$ merges, and the first flush in $\frac{\sizeratio - 1}{K_i} - 1$ merges as new flushes are eagerly compacted into the sorted run.
Therefore, the total count of merge operations for $K_i$ sorted runs on Level $i$ is
\begin{equation} \label{eq:merges-per-level}
    K_i \cdot \sum_{j=1}^{\frac{T-1}{K_i} - 1} j =
    (T - 1) \cdot \frac{(T - 1 - K_i)}{2K_i}.
\end{equation}
Given the total merges for Level $i$, we can now calculate the average number of merges a single write participates in.
First, we divide the total merges at Level $i$ from Equation~\eqref{eq:merges-per-level} by the number of flushes from Level $i - 1$ ($T - 1$) to receive an average merge count of $\frac{(T - 1 - K_i)}{2K_i}$.
Second, to account for the final full-level merge that occurs on the $T^{th}$ flush, we add 1 additional merge.
Therefore, the average number of merges, and subsequently I/Os, a single write participates in at Level $i$ is simply $\frac{T - 1 + K_i}{2K_i}$

To calculate the cost of a single insert, we need to divide the average number of merges every level by the number of entries per page, $B$. 
Additionally, as every compaction operation reads data at Level $i-1$ and writes to Level $i$, we model the potential asymmetry between reads and writes on the underlying storage device\footnote{Flash-based SSDs typically exhibit a read/write asymmetry, where writes are 2$\times$ to 10$\times$ more expensive than reads~\cite{Papon2021a}.} using $\asym$.
For example, a device for which a write operation is twice as expensive as a read operation has $\asym=2$.
When flushing the buffer, writes perform sequential I/Os as opposed to random I/Os, hence, we add $f_{seq}$ term to account for the cost of different I/O types.
Summing the average I/Os per level for all levels, the total I/O cost is captured by:

\begin{equation}\label{eq:write}
W(\configuration) =
    f_{seq} \cdot \frac{1 + \asym}{B}
    \cdot \sum_{i=1}^{L(\sizeratio)}
        \frac{\sizeratio - 1 + K_i}{2K_i}.
\end{equation}

\Paragraph{Total Expected Cost} 
The total expected operation cost, $\cost(\workload, \configuration)$, is computed by weighing the empty point lookup cost $Z_0(\configuration)$ from Equation~\eqref{eq:read}, the non-empty point lookup cost $Z(\configuration)$ from Equation~\eqref{eq:read-non-empty}, the range lookup cost $Q(\configuration)$ from Equation~\eqref{eq:range}, and the write cost $W(\configuration)$ from Equation~\eqref{eq:write} by their proportion in the workload represented by the terms $z_0$, $z$, $q$ and $w$ respectively (note that $z_0 + z_1 + q + w = 1$).

\subsection{Expressing LSM Data Layout Variants}

With the introduction of $K_i$, we can use our cost model to effectively describe the behavior of other common LSM designs.
For example, for a classic leveling compaction policy we set $\forall i, K_i = 1$.
This results in each level eagerly merging incoming compacted runs into a single run, which is equivalent to leveling.
Additionally, our cost model can easily describe other flexible LSM compaction behavior.
If we restrict all capacities before the last level to be equivalent (i.e., $K_1 = K_2 = ... = K_{L - 1}$ where $L$ is the last level), our cost model expresses the {\fluidmodel} design as described in {\dostmodel}~\cite{Dayan2018}.
With $K_{L} = 1$ and all $K_1 = K_2 = ... = K_{L - 1} = T - 1$, we have an equivalent cost model for {\llmodel}.
However, our proposed {\kmodel} is the most flexible, as each level has an independent limit on the number of runs.
Table~\ref{tab:lsm_variations} summarizes how {\kmodel} describes other common LSM variations:

\begin{table}[h]
    \caption{Variations of LSM data layouts}
    \label{tab:lsm_variations}
    \begin{center}
        \begin{tabular}[c]{ll}
           \toprule
                \multicolumn{1}{l}{\textbf{LSM layout}} &
                \multicolumn{1}{l}{\textbf{Setting}} \\
           \midrule
            {\fluidmodel}~\cite{Dayan2018} &
                $K_1 = ... = K_{L - 1}$ \\
            {\llmodel}~\cite{Dayan2018,Dayan2019} &
                $K_{L} = 1, K_i = T - 1,\ \forall i \neq L$ \\
            {\onelmodel}~\cite{Sarkar2021c} &
                $K_1 = T - 1, K_i = 1,\ \forall i \neq 1$ \\
            Tiering~\cite{Jagadish1997} &
                $K_i = T - 1,\ \forall i$ \\
            Leveling~\cite{ONeil1996} &
                $K_i = 1,\ \forall i$ \\
            {\kmodel} (\S\ref{sec:lsm-cost-model}) &
                All $K_i \in \mathbb{Z}$  \\
           \toprule
        \end{tabular}
    \end{center}
    \vspace{-0.25in}
\end{table}

\section{The Nominal Tuning Problem} \label{sec:tuning-lsm-trees}
In this section, we describe the classic tuning problem which involves finding the best configuration suited for a given workload without uncertainty.
We examine the problem definition, algorithms to efficiently compute optimal configurations, and compare configurations across various designs of LSM trees.

\subsection{{\nominal} Problem Definition}
Traditionally, designers focus on finding the configuration ${\configuration}^\ast$ that minimizes the total cost $\cost(\workload, \configuration^\ast)$, for a given fixed workload $\workload$.
We call this problem the {\nominal} problem, which is defined as follows:

\begin{problem}[{\nominal}]\label{problem:nominal}
Given a fixed workload $\workload$, find the LSM tree configuration
$\configuration_N$ such that
\begin{equation}
\label{eq:nominal_problem}
    \configuration_N =
        \argmin_{\configuration} \cost(\workload, \configuration).
\end{equation}
\end{problem}

\noindent
The problem described above captures the classic tuning paradigm of finding a system configuration that minimizes a cost model (describing I/O or latency) given a specific static workload and system environment.
Therefore, each LSM design described in Table~\ref{tab:lsm_variations} has a different {\nominal} problem based on the form of the cost function.
Prior tuning approaches either individually solve the {\nominal} problem solely for LSM data layouts~\cite{Dayan2018,Luo2020a} (e.g., tiering or leveling) or memory allocation~\cite{Dayan2017}, but not simultaneously for both.

\subsection{Solving a {\nominal} Problem}
To solve the {\nominal} problem, we utilize an off-the-shelf numerical solver.
We opt to use the Sequential Least Square Quadratic Solver (SLSQP) implemented in Python and packaged under the SciPy library~\cite{Virtanen2020}.
When choosing a data layout to optimize, we reduce the cost model to express the appropriate LSM design.

\Paragraph{Relaxing Integer Values}
Certain decision variables such as {\sizeratio} (size ratio) pose an issue as they are required to be integers as LSM trees cannot implement fractional size ratios.
To keep the problem feasible, we relax the integer constraint for such decision variables and opt to take the ceiling of any feasible solution before deploying the tuning.
In practice, this approach works well and leads to high-performance configurations.

\begin{figure}
    \begin{center}
        \includegraphics[width=\columnwidth]{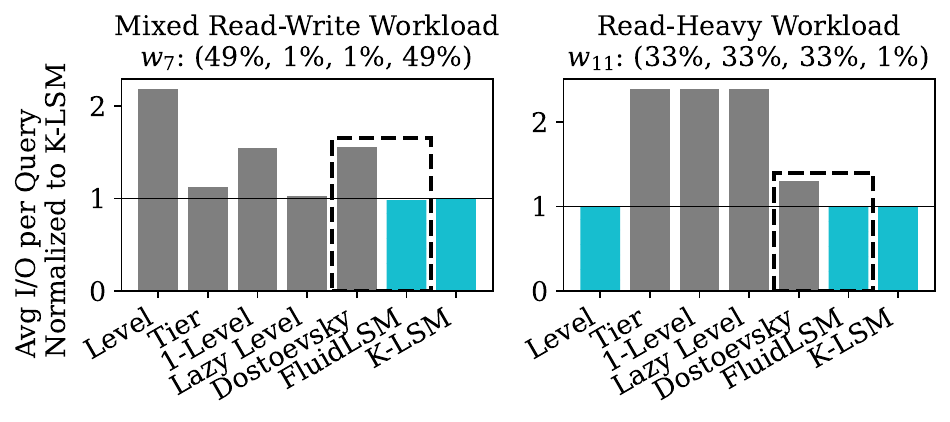}
    \end{center}
    \caption{
        Throughput of different designs for fixed workloads. Hatched-cyan
        indicates the best performance. Note {\dostmodel} uses the
        {\fluidmodel} design, but with fixed memory~\cite{Dayan2018}.
    }
    \label{fig:nominal-hist}
\end{figure}

\subsection{Comparison of LSM Strategies} \label{subsec:nominal-experiment}
In this section, we explore the optimal configurations for different designs described in Table~\ref{tab:lsm_variations} by solving the {\nominal} problem for each respective LSM design variation.
We compare average I/Os per query to analyze the performances of different flexible designs.

\Paragraph{Experiment Setup}
We adopt the following setting for system parameters: the database initially holds 10 billion entries, each of size 1 KB; a memory budget of 10 bits per element, or 10 GB in total divided among Bloom filter and write buffer; and a page size of 4 KB. 
It should be noted that the original {\dostmodel} strategy uses {\fluidmodel} as an LSM design with fixed memory allocation, and only optimizes for the maximum number of runs for the upper levels, the lowest level, and size ratio while fixing memory.
Therefore, while evaluating {\dostmodel} we fix {\mfilt} to 10 bits per entry and {\mbuf} to 2 MB as in~\cite{Dayan2018}.
For all other design variations, we solve a {\nominal} problem that optimizes memory and design while fixing other memory allocations such as fence pointers and the random access buffer.

\Paragraph{Flexible Performance}
Figure~\ref{fig:nominal-hist} shows the average I/O performance of various tuning configurations normalized to {\kmodel} design across different workloads. We experiment with a mixed read-write (${\workload}_7$) and a read-heavy (${\workload}_{11}$) workload from the uncertainty benchmark, which is presented in detail in Section \ref{sec:uncertainty-benchmark}.
Note that ${\workload}_7$ would traditionally lead a designer to focus on tiering as writes make up a large portion of the workload, while ${\workload}_{11}$ would suggest a leveling policy would be best.
When solving for more flexible designs---in this instance {\kmodel} and {\fluidmodel}---we observe that the optimizer always produces the best tunings.
Because $\workload_{11}$ is a read-heavy distribution, the optimal configuration has a leveling policy, which is reinforced by observing that the optimal {\kmodel} and {\fluidmodel} designs chosen are equivalent leveling.
For the balanced read-write workload $\workload_7$, we see that flexible designs outperform traditional designs as the optimizer finely tunes the capacity per level to accommodate both reads and writes.

\section{The Robust Tuning Problem}
\label{sec:robust-tuning}
In this section, we introduce the {\robustw} problem, a variation of the {\nominal} problem that takes into consideration uncertainty in the workload.
We give a precise definition of workload uncertainty and show how to compute high-performance configurations that minimize the expected cost of operation in the presence of this uncertainty.

\subsection{Robust Problem Definition}
The {\nominal} problem assumes perfect information about the workload before deploying the system.
For example, we may assume that the input vector $\workload$ represents the workload for which we optimize, while in practice, $\workload$ is simply an estimate of what an observed workload may look like.
Hence, the configuration obtained by solving Problem~\ref{problem:nominal} may result in variable performance if the observed workload upon deployment varies greatly from the expected workload.

We capture this uncertainty by reformulating Problem~\ref{problem:nominal} to take into account variability observed in the input workload.
Given an expected workload $\workload$, we introduce the notion of the \emph{uncertainty region} of $\workload$, which we denote by $\mathcal{U}_\workload$.

We can define the robust version of Problem~\ref{problem:nominal}, under the assumption that there is uncertainty in the input workload as follows:

\begin{problem}[{\robustw}]\label{problem:robustw} 
Given $\workload$ and uncertainty region $\mathcal{U}_\workload$ find the tuning configuration of the LSM tree $\configuration_R$ such that
\begin{eqnarray}
\label{eq:robust_workload_problem}
\configuration_R &=& \argmin_{\configuration} \cost(\obsworkload,
    \configuration) \nonumber\\
    \textrm{s.t.,}&& \obsworkload \in \mathcal{U}_{\workload}.
\end{eqnarray}
\end{problem}

\noindent
Note that the above problem definition intuitively states the following: it recognizes that the input workload $\workload$ will not be observed exactly, and it assumes that any workload in $\mathcal{U}_\workload$ is possible.
Then, it searches for the configuration $\configuration_\workload$ that is best for the \emph{worst-case} scenario among all those in $\mathcal{U}_\workload$. 

The challenge in solving {\robustw} is that one needs to explore all the workloads in the uncertainty region to solve the problem.
In the next section, we show that this is not necessary.
In fact, by appropriately rewriting the problem definition we show that we can solve Problem~\ref{problem:robustw} in polynomial time.

% =============================================================================
\subsection{Solving The {\robustw} Problem}
\label{sec:algo-for-robust-tuning}
% =============================================================================

In this section, we discuss our solutions to the {\robustw} problem.
On a high level, the solution strategy is the following: first, we express the objective of the problem (as expressed in Equation~\eqref{eq:robust_workload_problem}) as a standard continuous optimization problem.  
We then take the \emph{dual} of this problem and use existing results in robust optimization to show: $(i)$ the duality gap between the primal and the dual is zero, and $(ii)$ the dual problem is solvable in polynomial time.
Thus, the dual solution can be translated into the optimal solution for the primal, i.e., the original {\robustw} problem.
The specifics of the methodology are described below:

\begin{figure}[t]
    \centering
    \begin{minipage}[b]{0.49\columnwidth}
        \includegraphics[width=\columnwidth]{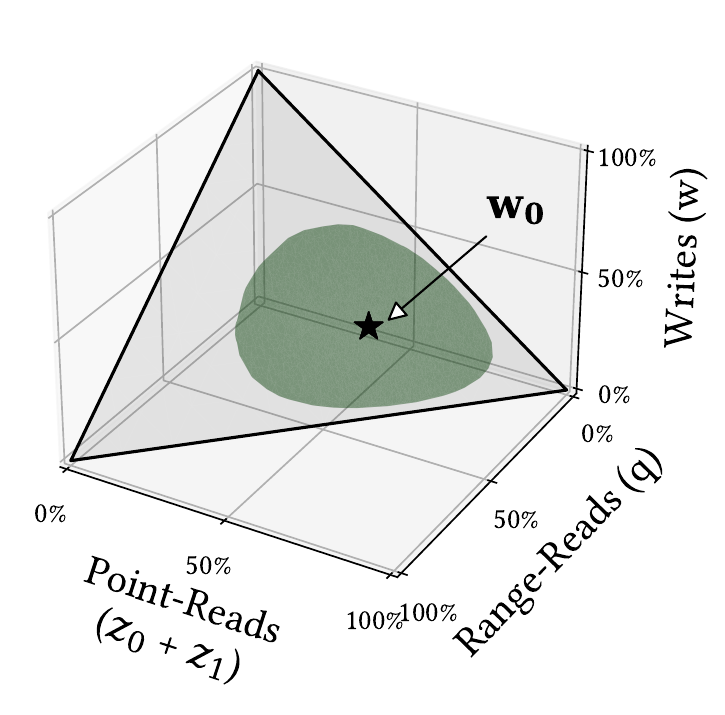}
    \end{minipage}
    \hfill
    \begin{minipage}[b]{0.49\columnwidth}
        \includegraphics[width=\columnwidth]{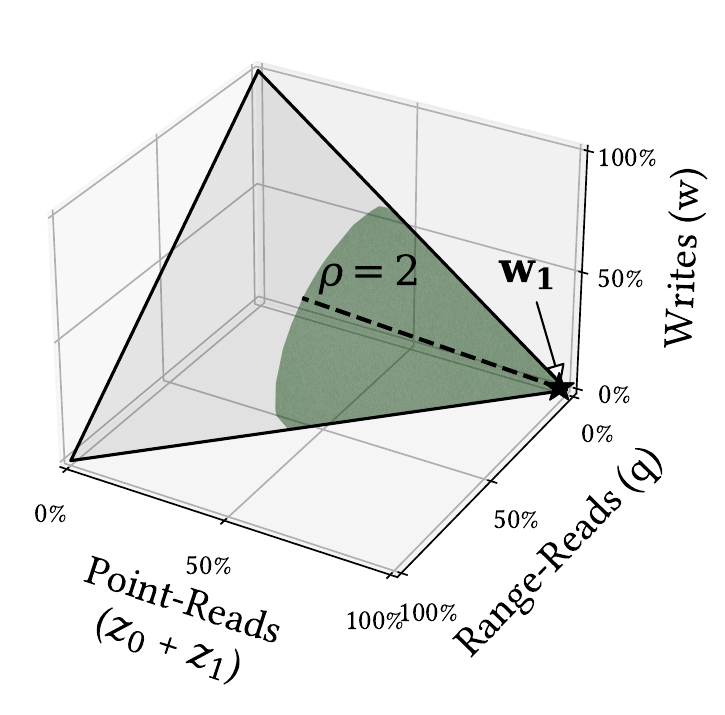}
    \end{minipage}
    \caption{Workload uncertainty neighborhoods ($\mathcal{U}_\workload$),
        denoted by the green shaded region, for two different expected workloads
    ($\workload$) and $\rho$.}
    \label{fig:kl_div_plane}
\end{figure}

\Paragraph{Defining the Uncertainty Region $\mathcal{U}_\workload$}
Recall that $\workload$ is a probability vector, i.e., $\workload^\intercal\columnvec = 1$. Thus, in order to define the uncertainty region $\mathcal{U}_\workload$, we use the Kullback-Leibler (KL) divergence function~\cite{Kullback1951a} defined as follows:
\begin{definition}\label{def:KL}
The KL-divergence distance between two probability distributions $\vec{p} = (p_1, \cdots, p_m)^\intercal \geq 0$ and $\vec{q} = (q_1, \cdots, q_m)^\intercal \geq 0$  is defined as,
\begin{equation*}
    I_{KL}(\vec{p}, \vec{q}) = \sum_{i=1}^{m}p_i \log\bigg(\frac{p_i}{q_i}\bigg).
\end{equation*}
\end{definition}
\noindent Since our workloads are represented as probability distributions, the KL-divergence is the most natural choice of distance between them.
One could use $L_p$ norms instead.
However, calculating the $L_p$ norm between workloads requires a summation of the $p^{\textrm{th}}$ power of differences in probabilities, which are extremely small values, and are not meaningful in this setting.

Using the KL-divergence we can now formalize the definition of the uncertainty region around an expected workload {\workload} as follows,
\begin{equation}
\label{eq:workloaduncertaintyregion}
    \mathcal{U}_{\workload}^\rho =\{
        \obsworkload \in \mathbb{R}^4\ |\ 
        \obsworkload \geq 0,
        \obsworkload^\intercal\columnvec = 1,
        I_{KL}(\obsworkload, \workload) \leq \rho
\}.
\end{equation}

\noindent Here, $\rho$ determines the maximum KL-divergence that is allowed between any workload $\obsworkload$ in the uncertainty region and the input expected workload $\workload$.
Note that the definition of the uncertainty region takes as input the parameter $\rho$, which intuitively defines the neighborhood around the expected workload.
Figure \ref{fig:kl_div_plane} shows an example of the uncertainty region for $\rho = 0.2$ and expected workload $\workload_0 = (25\%, 25\%, 25\%, 25\%)$, and for $\rho = 2$ and expected workload $\workload_1 = (97\%, 1\%, 1\%, 1\%)$.
For this visualization, we combined the two types of read queries (empty and non-empty) onto one axis.
Note that the shape of the uncertainty region is defined by the expected workload, the value of $\rho$, and the fact that all workloads are restricted to be probability distributions. 
In terms of notation, $\rho$ is required for defining the uncertainty region $\mathcal{U}_\workload^\rho$.
However, we drop the superscript notation unless required for context.

\Paragraph{Rewriting of the ROBUST TUNING Problem (Primal)}
Using the above definition of the workload uncertainty region $\mathcal{U}_\workload^\rho$, we are now ready to proceed to the solution of the {\robustw} problem.
For a given $\rho$, the problem definition, as captured by Equation~\eqref{eq:robust_workload_problem}, can be rewritten as follows:
\begin{equation}\label{eq:robust_tuning_problem1}
    \min_{\configuration} \max_{\obsworkload \in \mathcal{U}_{\workload}^\rho}
    \obsworkload^\intercal \costvec(\configuration).
\end{equation}

\noindent This rewrite captures the intuition that the optimization is done over the \emph{worst-case} scenario across all the workloads in the uncertainty region $\mathcal{U}_\workload$.
Equation~\eqref{eq:robust_tuning_problem1} can be rewritten by introducing an additional variable $\beta\in\mathbb{R}$, as follows:
\begin{eqnarray}
\label{eq:robust_counterpart_workload_problem}
\min_{\beta, \configuration}&\beta& \nonumber\\
    \textrm{s.t.,}&\obsworkload^\intercal\costvec(\configuration) \leq \beta
                  &\forall  \obsworkload \in \mathcal{U}_{\workload}.
\end{eqnarray}
This reformulation allows us to remove the $\min\max$ term in the objective from Equation~\eqref{eq:robust_tuning_problem1}.
The constraint in Equation~\eqref{eq:robust_counterpart_workload_problem} can be equivalently expressed as,
\begin{eqnarray*}
\label{eq:inner_optimization}
    \beta &\geq&
    \max_{\obsworkload}\big\{\obsworkload^\intercal\costvec(\configuration) | \obsworkload \in
    \mathcal{U}_{\workload}\big\}\nonumber\\
    &=&\max_{\obsworkload \geq
0}\bigg\{\obsworkload^\intercal\costvec(\configuration)\bigg|\obsworkload^\intercal\columnvec
= 1, \sum_{i=1}^{m} \hat{w}_i \log\bigg(\frac{\hat{w}_i}{w_i}\bigg) \leq \rho\bigg\}.
\end{eqnarray*}
Finally, the Lagrange function for the optimization on the right-hand side of the above equation is:
\begin{equation*}
    \mathcal{L}(\obsworkload, \lambda, \eta) =
    \obsworkload^\intercal\costvec(\configuration) + \rho \lambda - \lambda
    \sum_{i=1}^{m}\hat{w}_i \log\bigg(\frac{\hat{w}_i}{w_i}\bigg) + \eta(1 -
    \obsworkload^\intercal \columnvec),
\end{equation*}
where $\lambda$ and $\eta$ are the Lagrangian variables. 

\Paragraph{Formulating the Dual Problem}
We can now express the dual objective as,
\begin{equation}\label{eq:dual}
    g(\lambda, \eta) = \max_{\obsworkload \geq 0}\mathcal{L}(\obsworkload,
    \lambda, \eta),
\end{equation}
which we need to \emph{minimize}.

Now we borrow the following result from~\cite{Ben-Tal2013},
\begin{lemma}[\cite{Ben-Tal2013}]
A configuration {\configuration} is the optimal solution to the {\robustw} problem if and only if $\min_{\eta, \lambda \geq 0}g(\lambda, \eta) \leq \beta$ where the minimum is attained for some value of $\lambda \geq 0$.
\end{lemma}
In other words, minimizing the dual objective $g(\lambda, \eta)$ -- as expressed in Equation~\eqref{eq:dual} -- will lead to the optimal solution for the {\robustw} problem.

\Paragraph{Solving the Dual Optimization Problem Optimally} 
Formulating the dual problem and using the results of Ben-Tal {\etal}~\cite{Ben-Tal2013}, we have shown that the dual solution leads to the optimal solution for the {\robustw} problem.
Moreover, we can obtain the optimal solution to the original {\robustw} problem in polynomial time, a consequence of the tractability of the dual objective.

To solve the dual problem, we first simplify the dual objective $g(\lambda,\eta)$ so that it takes the following form:
\begin{equation}\label{eq:final_dual_rewrite}
 g(\lambda,\eta) = \eta + \rho \lambda + \lambda
    \sum_{i=1}^{k}w_i\phi_{KL}^* \bigg(\frac{\costvec_i(\configuration) - \eta}{\lambda}\bigg).
\end{equation}

\noindent In Equation~\eqref{eq:final_dual_rewrite}, $\phi_{KL}^*(.)$ denotes the conjugate of KL-divergence function and $\costvec_i$ corresponds to the $i$-th dimension of the cost vector $\costvec(\configuration)$ as defined in Section~\ref{sec:notation} -- clearly in this case $k=4$ as we have $4$ types of queries in our workload.
Results of Ben-Tal {\etal}~\cite{Ben-Tal2013} show that minimizing the dual function as described in Equation~\eqref{eq:final_dual_rewrite} is a convex optimization problem, and it can be solved optimally in polynomial time if and only if the cost function $\costvec(\configuration)$ is convex in all its dimensions.

In our case, the cost function for the range queries is not convex w.r.t. size ratio {\sizeratio} for the tiering policy.
However, on account of its smooth non-decreasing form, we are still able to find the global minimum solution for 
\begin{eqnarray}
\min_{\configuration, \lambda \geq 0, \eta}
\bigg\{\eta + \rho \lambda + \lambda
    \sum_{i=1}^{m}w_i\phi_{KL}^* \bigg(\frac{c_i(\configuration) -
\eta}{\lambda}\bigg)\bigg\}.
\end{eqnarray}
This minimization problem can be solved using the Sequential Least Squares Quadratic Programming solver (SLSQP) included in the popular Python optimization library SciPy~\cite{Virtanen2020}.
Solving this problem outputs the values of the Lagrangian variables $\lambda$ and $\eta$ and most importantly the configuration $\configuration$ that optimizes the objective of the {\robustw} problem -- for input $\rho$.
In terms of running time, the SLSQP solver outputs a robust tuning configuration for a given input in less than a second.

\Paragraph{Finding a Value for $\rho$}
Since $\rho$ is a robust tuning parameter, we also provide a few heuristics for setting it.
In the presence of historically observed workloads, a DBA may calculate $\rho$ using the following definition: that is, $\rho$ is set to be the largest KL-divergence between any observed workload and the corresponding workload average as described in Algorithm~\ref{alg:historical_rho}.
If the DBA does not have information about past workloads, they may provide ranges for each query type; then, can sample workloads within those ranges and then calculate $\rho$ using the definition above to find an appropriate value.
DBAs may instead provide two workloads, one that is expected during a normal observation period, and another off-period or unlikely workload.
In this case, The KL-divergence between these two workloads can be used as $\rho$.

\begin{algorithm}
\DontPrintSemicolon
\KwIn{Set of historical workloads $\mathcal{W} = \{{\workload_1}, {\workload_2}, ..., {\workload_n}\}$}
    $\bar{\workload} \leftarrow \frac{1}{n} \cdot \sum\limits_{\workload_i \in \mathcal{W}} \workload_i$\;
    $\Return\ \argmax\limits_{\workload_i \in \mathcal{W}} I_{KL}(\workload_i, \bar{\workload})$
\caption{Calculating $\rho$ from historical workloads}
\label{alg:historical_rho}
\end{algorithm}

% =============================================================================
\section{Uncertainty Benchmark}
\label{sec:uncertainty-benchmark}
% =============================================================================

In this section, we describe the uncertainty benchmark that we use to evaluate the \Endure, both analytically using the cost models, and empirically using RocksDB.
It consists of two primary components: (1) \emph{Expected workloads} and, (2) \emph{Benchmark set of sampled workloads}, described below.

\Paragraph{Expected Workloads}
We create robust tuning configurations for 15 expected workloads encompassing different proportions of query types.
We catalog them into \emph{uniform}, \emph{unimodal}, \emph{bimodal}, and \emph{trimodal} categories based on the dominant query types.
While this breakdown of dominant queries is similar to benchmarks such as YCSB, we provide a more comprehensive coverage of potential workloads.
A minimum of 1\% of each query type is always included in every expected workload to ensure a finite KL-divergence.
A complete list of all expected workloads is in Table~\ref{tab:expected-workloads}.

\begin{table}[ht]
   \centering%\small
   \caption{Tested expected workloads.}
   \label{tab:expected-workloads}	
   \begin{tabular}{c cccc l}
   \toprule
   Index & \multicolumn{4}{c}{$(\emptylookup, \nonemptylookup, \range, \update)$} & Type \\
   \toprule
   0 & 25\% & 25\% & 25\% & 25\% & \textbf{Uniform} \\
   \midrule
   1 & 97\% & 1\% & 1\% & 1\% & \textbf{Unimodal}\\
   2 & 1\% & 97\% & 1\% & 1\% & \\
   3 & 1\% & 1\% & 97\% & 1\% & \\
   4 & 1\% & 1\% & 1\% & 97\% & \\
   \midrule
   5 & 49\% & 49\% & 1\% & 1\% & \textbf{Bimodal}\\
   6 & 49\% & 1\% & 49\% & 1\% & \\
   7 & 49\% & 1\% & 1\% & 49\% & \\
   8 & 1\% & 49\% & 49\% & 1\% & \\
   9 & 1\% & 49\% & 1\% & 49\% & \\
   10 & 1\% & 1\% & 49\% & 49\% & \\
   \midrule
   11 & 33\% & 33\% & 33\% & 1\% & \textbf{Trimodal}\\
   12 & 33\% & 33\% & 1\% & 33\% & \\
   13 & 33\% & 1\% & 33\% & 33\% & \\
   14 & 1\% & 33\% & 33\% & 33\% & \\
   \bottomrule
   \end{tabular}
\end{table}

\Paragraph{Benchmark Set of Sampled Workloads}
We use the benchmark set of 10K workloads {\benchmark} as a \emph{test} dataset over which to evaluate the tuning configurations.
These configurations are generated as follows: first, we independently sample the number of queries corresponding to each query type uniformly at random from a range $(0, 10000)$ to obtain a $4$-tuple of query counts.
Next, we divide the individual query counts by the total number of queries in the tuple to obtain a random workload that is added to the benchmark set.
We use the actual query counts during the system experimentation where we execute individual queries on the database.

This type of workload breakdown can commonly be seen in LSM trees as shown in a survey of workloads in Facebook's pipeline~\cite{Cao2020}.
The authors report that ZippyDB, a distributed KV store that uses RocksDB, experiences workloads with 78\% gets, 19\% writes, and 3\% range reads.
This breakdown is similar to workload $11$, and the exact workload is in the benchmark set {\benchmark}.

Note that while the same {\benchmark} is used to evaluate different tunings, it represents a different distribution of KL-divergences for the corresponding expected workload associated with each tuning.
In the next two sections, we use our uncertainty benchmark to demonstrate that tuning with {\Endure} achieves significant performance improvement using both a model-based analysis (Section~\ref{sec:model-evaluation}), and an experimental study (Section~\ref{sec:system-evaluation}).

\section{Model-Based Evaluation}
\label{sec:model-evaluation}
% =============================================================================

We now present our detailed model-based study of {\Endure} that uses more than 10000 different noisy workloads for all 15 expected workloads, showing performance benefit of up to $5\times$.
For brevity, when we provide a nominal tuning we are referring to the solution for the {\nominal} problem with tiering and leveling as the two design choices.
Similarly, we use {\Endure} and the robust tuning interchangeably to refer to the solution of the {\robustw} problem which chooses between tiering and leveling.
We show that {\Endure} perfectly matches the nominal tuning when there is no uncertainty (i.e., when the observed workload always matches the expected one) and we pass this information to the robust tuner.
Further, we provide recommendations on how to select uncertainty parameters. 

\subsection{Evaluation Metrics}
In this section, we provide definitions of metrics used to evaluate the performance of tunings.

\Paragraph{Normalized Delta Throughput ($\Delta$)}
Defining throughput as the reciprocal of the cost of executing a workload, we measure the normalized delta throughput of a configuration {$\configuration_2$} w.r.t. another configuration {$\configuration_1$} for a given workload {\workload} as follows, 
\[
    \Delta_{\workload}(\configuration_1, \configuration_2) = 
    \frac{\nicefrac{1}{C(\workload, \configuration_2)} -
        \nicefrac{1}{C(\workload, \configuration_1)}}
        {\nicefrac{1}{C(\workload, \configuration_1)}}.
\]
$\Delta_{\workload}(\configuration_1, \configuration_2) > 0$ implies that {$\configuration_2$} outperforms {$\configuration_1$} when executing a workload {\workload} and vice versa when $\Delta_{\workload}(\configuration_1, \configuration_2) < 0$.

\Paragraph{Throughput Range ($\Theta$)}
While normalized delta throughput compares two different tunings, we use the throughput range to evaluate an individual tuning ${\configuration}$ w.r.t. the benchmark set {\benchmark} as follows,
\[
    \Theta_{\mathcal{B}}(\configuration) = 
    \max_{\workload_0, \workload_1 \in \mathcal{B}}
        \bigg(\frac{1}{C(\workload_0, \configuration)} 
        - \frac{1}{C(\workload_1, \configuration)}\bigg).
\]
$\Theta_{\benchmark}(\configuration)$ intuitively captures the best and the worst-case outcomes of the tuning {\configuration}.
A smaller value of this metric implies higher consistency in performance.

\begin{figure}
    \centering
    \includegraphics[width=\columnwidth]{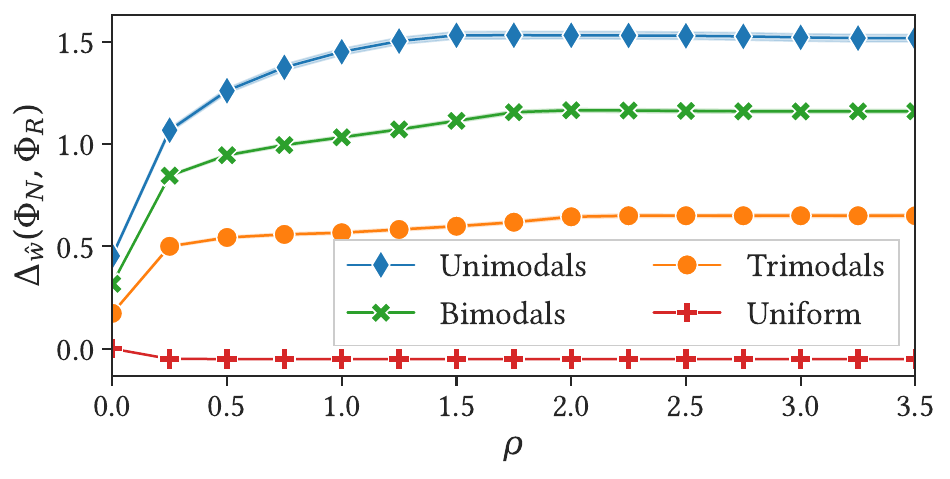}
    \caption{
        Average delta throughput $\Delta_{\hat{w}}(\Phi_{N}, \Phi_{R})$ for each category of expected workload.
    }
    \label{fig:delta_throughput_workload_type}
\end{figure}

\begin{figure*}
    \centering
    \includegraphics[width=\textwidth]{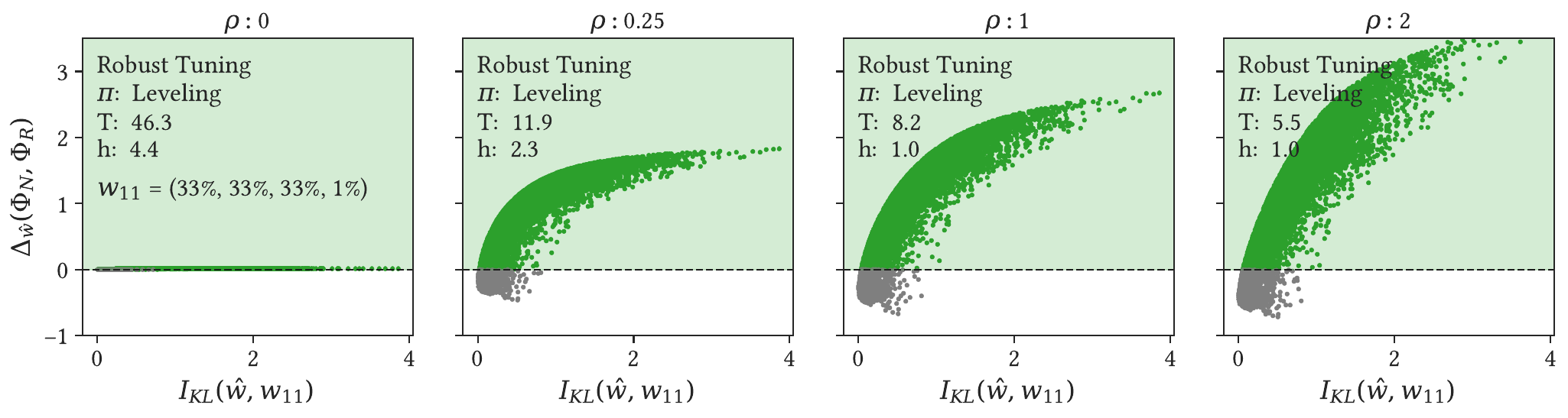}
    \caption{
        Impact of $\rho$ on normalized delta throughput $\Delta_{\obsworkload}(\configuration_{N}, \configuration_{R})$ for tunings with expected workload $\workload_{11}$.
    }
    \label{fig:scatterplot_evolution_rho}
\end{figure*}

\begin{figure*}[tbp]
    \centering
    \captionsetup{aboveskip=0.5em}
    \begin{subfigure}[h]{0.74\textwidth}
        \includegraphics[scale=0.5]{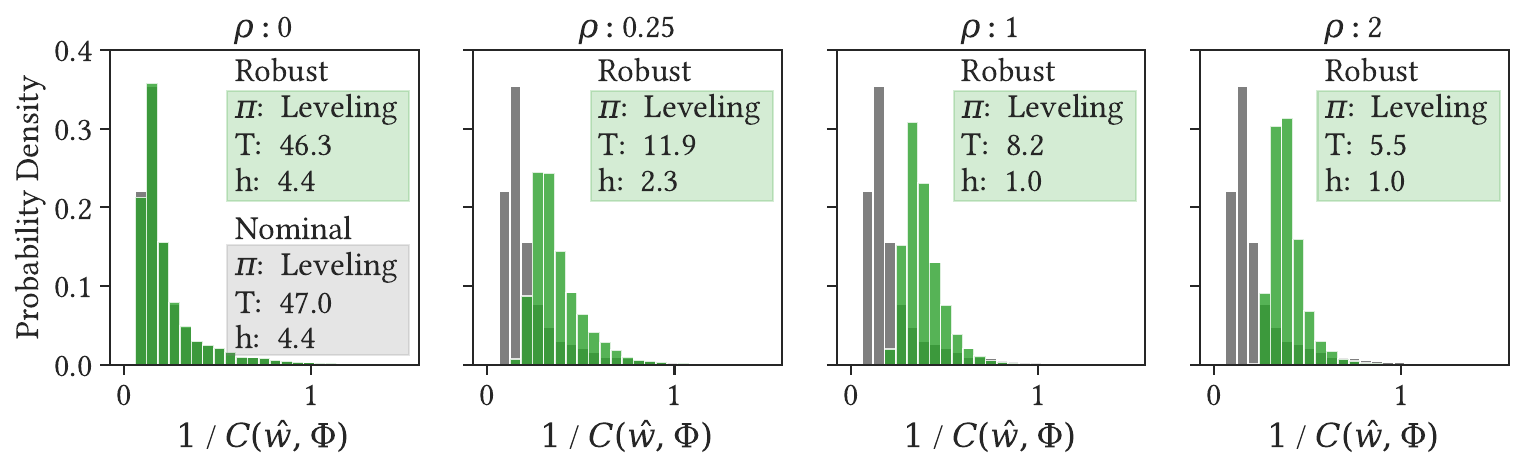}
        \caption{
            Histograms of throughput $1 / C(\obsworkload, \configuration)$ for tunings with expected workload $\workload_{11}$.
        }
        \label{fig:overlapping_histogram}
    \end{subfigure}
    \begin{subfigure}[h]{0.24\textwidth} 
        \includegraphics[scale=0.5]{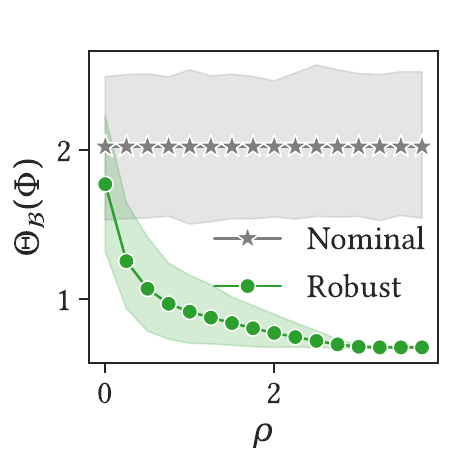}
        \caption{Throughput range $\Theta_{\mathcal{B}}(\configuration)$}
        \label{fig:throughput_range_evolution}
    \end{subfigure}
    \caption{Impact of $\rho$ on throughput.}
    \label{fig:rho_throughput_range_impact}
\end{figure*}

\subsection{Experiment Design}
To evaluate the performance of our proposed robust tuning approach, we design a large-scale experiment comparing different tunings over the sampled workloads in {\benchmark} using the analytical cost model.
For each of the expected workloads in Table \ref{tab:expected-workloads}, we obtain a single nominal tuning configuration (${\configuration_N}$) by solving the {\nominal} problem. 
For 15 different values of $\rho$ in the range (0.0, 4.0) with a step size of 0.25, we obtain a set of robust tuning configurations ($\configuration_R$) by solving the {\robustw} problem.
Finally, we individually compare each of the robust tunings with the nominal over the 10,000 workloads in {\benchmark} to obtain over 2 million comparisons.
While computing the costs, we assume that the database contains 10 billion entries each of size 1 KB.
The analysis presented in the following sections assumes a total available memory of 10 GB.
For brevity, we present representative results corresponding to individual expected workloads and specific system parameters.
We primarily focus on two workloads from Table~\ref{tab:expected-workloads}, $\workload_{7}$ which is a mixed read-write workload, and $\workload_{11}$, which is a read-heavy workload.
However, we exhaustively confirmed that changing these parameters does not qualitatively affect the outcomes of our experiment.

\subsection{Results}
\label{sec:model-results}
Here, we present an analysis of the comparisons between the robust and the nominal tuning configurations.
Using an off-the-shelf global minimizer from the popular Python optimization library SciPy~\cite{Virtanen2020}, we obtain both nominal and robust tunings with the runtime for the above experiment being less than 10 minutes.

\Paragraph{Comparison of Tunings}
First, we address the question -- \textit{is it beneficial to adopt robust tunings relative to the nominal tunings?}
Intuitively, it should be clear that the performance of nominally tuned databases would degrade when the workloads being executed on the database are significantly different from the expected workloads used for tuning.
In Figure~\ref{fig:delta_throughput_workload_type}, we present performance comparisons between the robust and the nominal tunings for different values of uncertainty parameter $\rho$.
We observe that robust tunings provide substantial benefit in terms of normalized delta throughput for \emph{unimodal}, \emph{bimodal}, and \emph{trimodal} workloads.
The normalized delta throughput $\Delta_\obsworkload(\configuration_N, \configuration_R)$ shows over 95\% improvement on average over all $\obsworkload \in \benchmark$ for robust tunings with $\rho \geq 0.5$, when the expected workload used during tuning belongs to one of these categories.
For \emph{uniform} expected workload, we observe that the nominal tuning outperforms the robust tuning by a modest 5\%.

Intuitively, \emph{unbalanced} workloads result in overfit nominal tunings.
Hence, even small variations in the observed workload can lead to significant degradation in the throughput of such nominally tuned databases. 
On the other hand, robust tunings by their very nature take into account such variations and comprehensively outperform the nominal tunings.
In the case of the uniform expected workload $\workload_0$, a low value of $\rho$ covers a larger area of possible workloads than that same value would in a different workload as evident in Figure~\ref{fig:kl_div_plane}.
In this case, when tuned for high values of $\rho$, the robust tunings are unrealistically pessimistic and lose performance relative to the nominal tuning.

\Paragraph{Impact of Tuning Parameter $\rho$}
Next, we address the question -- \emph{how does the uncertainty tuning parameter $\rho$ impact the performance of the robust tunings?}
In Figure~\ref{fig:scatterplot_evolution_rho}, we take a deep dive into the performance of robust tunings for an individual expected workload for different values of $\rho$.
We observe that the robust tunings for $\rho=0$ i.e., zero uncertainty, are very close to the nominal tunings.
As the value of $\rho$ increases, its performance advantage over the nominal tuning for the observed workloads with higher KL-divergence w.r.t. expected workload increases.
Furthermore, the robustness of such configurations have logically sound explanations.
The expected workload in Figure~\ref{fig:scatterplot_evolution_rho} consists of just 1\% writes.
Hence, for low values of $\rho$, the robust tuning has a higher size ratio leading to shallower LSM trees to achieve good read performance.
For higher values of $\rho$, the robust tunings anticipate an increasing percentage of write queries and hence limit the size ratio to achieve higher throughput.

In Figure~\ref{fig:rho_throughput_range_impact}, we show the impact of tuning parameter $\rho$ on the throughput range.
In Figure~\ref{fig:overlapping_histogram} we plot a histogram of the nominal and robust throughputs for workload $\workload_{11}$. 
As the value of $\rho$ increases, the interval size between the lowest and the highest throughputs for the robust tunings consistently decreases.
We provide further evidence of this phenomenon in Figure~\ref{fig:throughput_range_evolution}, by plotting the decreasing throughput range $\Theta_\benchmark(\configuration_R)$  averaged across all the expected workloads. 
Thus, robust tunings not only provide a higher average throughput for all $\obsworkload \in \benchmark$, but they have a more consistent performance (lower variance) compared to the nominal tunings.

\begin{figure}
    \centering
    \includegraphics[width=\columnwidth]{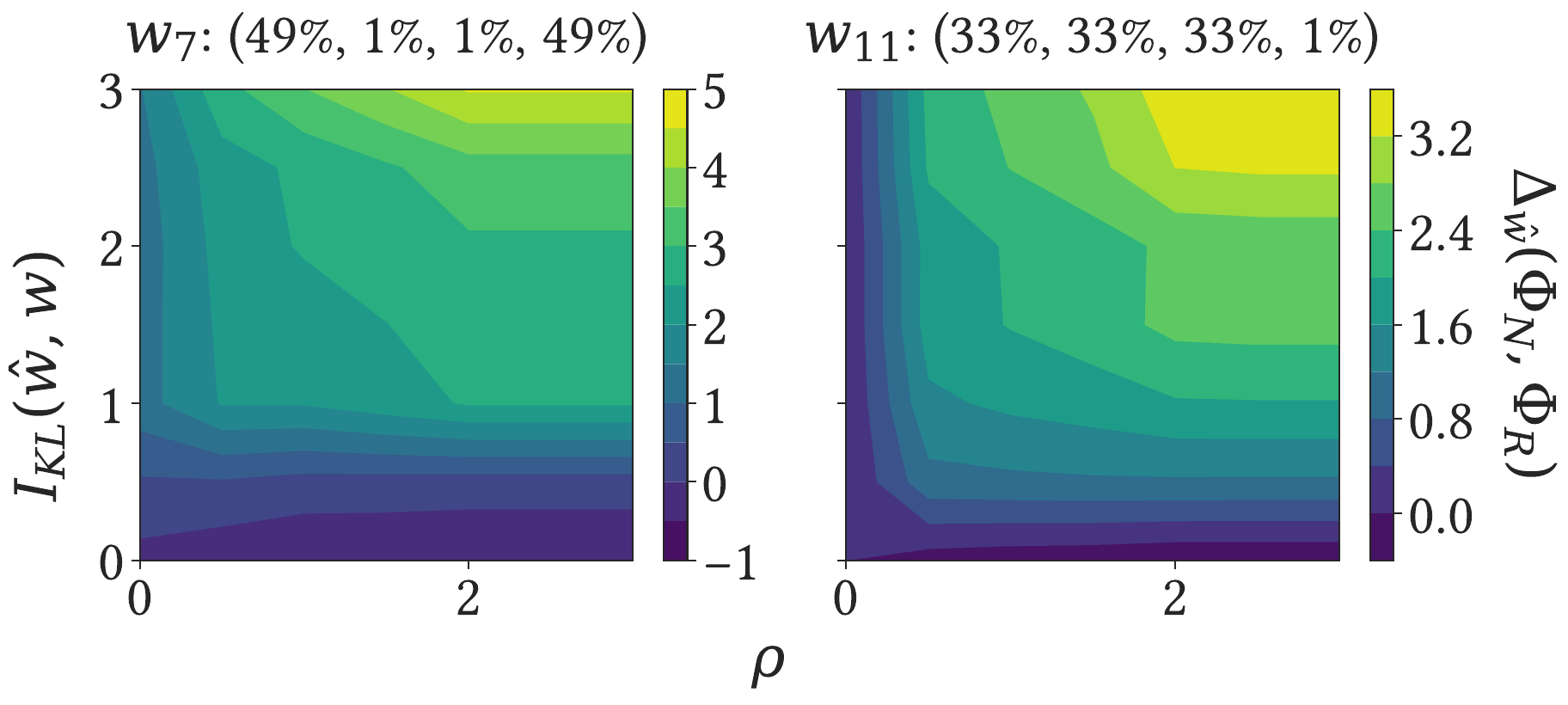}
    \caption{
        Effect on delta throughputs $\Delta_{\obsworkload}(\configuration_{N}, \configuration_{R})$ on selection of $\rho$ vs $I_{KL}(\obsworkload, \workload)$.
    }
    \label{fig:rho_vs_rho_hat}
\end{figure}

\Paragraph{Choice of $\rho$} Now, we address the question -- 
\emph{What is the appropriate choice for the value of uncertainty parameter $\rho$?}
In Figure~\ref{fig:rho_vs_rho_hat}, we explore the relationship between $\rho$ and the KL-divergence $I_{KL}(\obsworkload, \workload)$ for $\obsworkload \in \benchmark$, by making a contour plot of the corresponding normalized delta throughput $\Delta_{\obsworkload}(\configuration_N, \configuration_R)$.
We confirm our intuition that nominal tunings compare favorably with our proposed robust tunings only in two scenarios: (1) when the observed workloads are extremely similar to the expected workload (close to zero observed uncertainty), and (2) when the robust tunings assume extremely low uncertainty with $\rho < 0.2$ while the observed variation is higher.
Based on this evidence, we propose the following rule of thumb: the maximum KL-divergence between any two pairs of observed workloads is a reasonable value of $\rho$  in practice.

\begin{figure}
    \centering
    \includegraphics[width=\columnwidth]{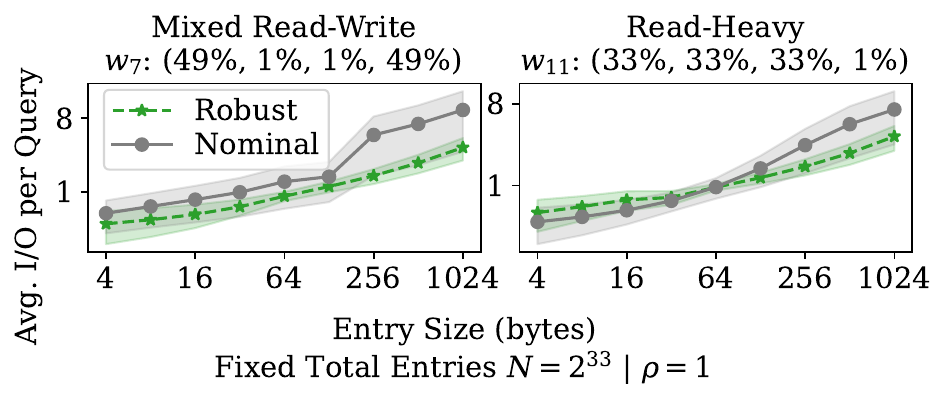}
    \caption{
        Tuning performance sensitivity to entry size.
    }
    \label{fig:entry_size_sensitivity}
\end{figure}

\Paragraph{Sensitivity Analysis w.r.t. Entry Size}
Lastly, we take a look at the expected tuning performance w.r.t to system settings.
Figure~\ref{fig:entry_size_sensitivity} shows average I/O, or single logical page accesses, per query over different entry sizes with the standard deviation highlighted around each line.
Each data point corresponds to the average I/O per query for the optimal tuning for all workloads $\hat{\workload} \in \benchmark$.
For our mixed read-write workload, we see that the {\Endure} always performs better than the nominal tuning regardless of the entry size.
When we tune with a read-heavy workload as the expected input, we observe that for lower entry sizes the nominal tuning produces a better tuning, however, at larger entry sizes {\Endure} outperforms its nominal counterpart.
Because the total number of entries is fixed, lower entry sizes cause the physical size of the database to be relatively small w.r.t. to the available memory budget.
Hence, we observe the allocation between {\mfilt} and {\mbuf} does not play a major role in performance as the tree can be made relatively shallow.
However, as the size of the database starts to increase and the memory budget becomes a smaller fraction of the database size, we observe the allocation between memory plays a larger role.
This implies proper robust tunings play a larger role in constrained environments, where the available memory is a small fraction of the total database size.

% =============================================================================
\section{System-Based Evaluation}
\label{sec:system-evaluation}
% =============================================================================
In this section, we deploy {\Endure} as the tuner of the state-of-the-art LSM-based engine RocksDB, and we show that RocksDB achieves up to 90\% lower workload latency in the presence of uncertainty.
We further show that the tuning cost is negligible, and the effectiveness of {\Endure} is not affected by data size.

\subsection{Experimental Setup \& Measurements}

Our server is configured with two Intel Xeon Gold 6230 processors, 384 GB of main memory, a 1 TB Dell P4510 NVMe drive, CentOS 7.9.2009, and a default page size of 4 KB.
We use Facebook's RocksDB, a popular LSM tree-based storage system, to evaluate our approach~\cite{FacebookRocksDB}.
While RocksDB provides implementations of leveling and tiering policies, the system implements micro-optimizations not common across all LSM tree-based storage engines.
Therefore, we use RocksDB's event hooks to implement both classic leveling and tiering policies to benchmark the common compaction strategies.
For default RocksDB comparisons, we set a custom policy hook to match the default compaction policy of leveling.
Additionally, RocksDB does not toggle on Bloom filters by default. 
In the interest of fair comparison, we add Bloom filters with the bits per element set to $10$.
Following the Monkey memory allocation scheme~\cite{Dayan2017}, we allocate different bits per element for Bloom filters per level.
We note that turning off direct I/O improves read performance for any tuning deployed to RocksDB. 
However, to obtain an accurate count of block accesses we instead enable direct I/Os for both queries and compaction and disable the block cache.
To obtain detailed insights about accesses, we present our findings with direct I/Os, however, our qualitative results remain unchanged with direct I/Os turned off.
Lastly, portions of memory reserved for the fence pointer and max read buffer are fixed to their default values before performing any tuning for buffer size and Bloom filter memory.

\begin{figure}[t]
    \begin{center}
        \includegraphics[width=\columnwidth]{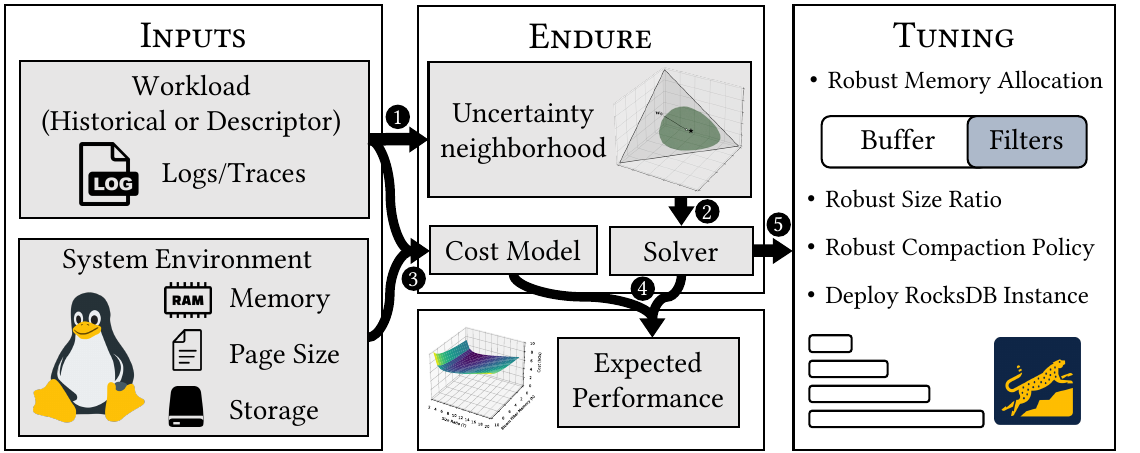}
    \end{center}
    \caption{
        (1) Workload information is provided to {\Endure}, establishing an uncertainty neighborhood centered on the expected workload.
        (2) This description of workload uncertainty is integrated into the solver.
        (3) The cost model receives both the workload and system parameter details.
        (4) Using the robust tuning from the solver and the cost model, an expected performance profile is generated.
        (5) The robust tuning is then deployed onto RocksDB.
    }
    \label{fig:endure_pipeline}
\end{figure}

\Paragraph{\textsc{Endure}'s Pipeline}
Figure~\ref{fig:endure_pipeline} shows the workflow used for {\Endure} and the following experiments.
A workload descriptor (expected workload and uncertainty value $\rho$) is provided to {\Endure} to create an uncertainty neighborhood centered around an expected workload.
This description of workload uncertainty is then incorporated into the solver.
In combination with the cost model, which uses the workload and system parameters as inputs, {\Endure} outputs an expected performance profile and a robust tuning over various workloads in the uncertainty neighborhood.
{\Endure} then deploys the robust tuning on a RocksDB instance where we execute workloads to measure performance.

\Paragraph{Empirical Measurements}
We use the internal RocksDB statistics module to measure the number of logical block accesses during reads, bytes flushed during writes, and bytes read and written in compactions.
The number of logical blocks accessed during writes is calculated by dividing the number of bytes reported by the default page size.
To estimate the amortized cost of writes, we compute the I/Os from compactions across all workloads of a session and redistribute them across write queries.
Our approach of measuring average I/Os per query allows us to compare the effects of different tuning configurations, while simultaneously minimizing the effects of extraneous factors on performance.

\begin{figure*}[t]
    \centering
    \includegraphics[width=\linewidth]{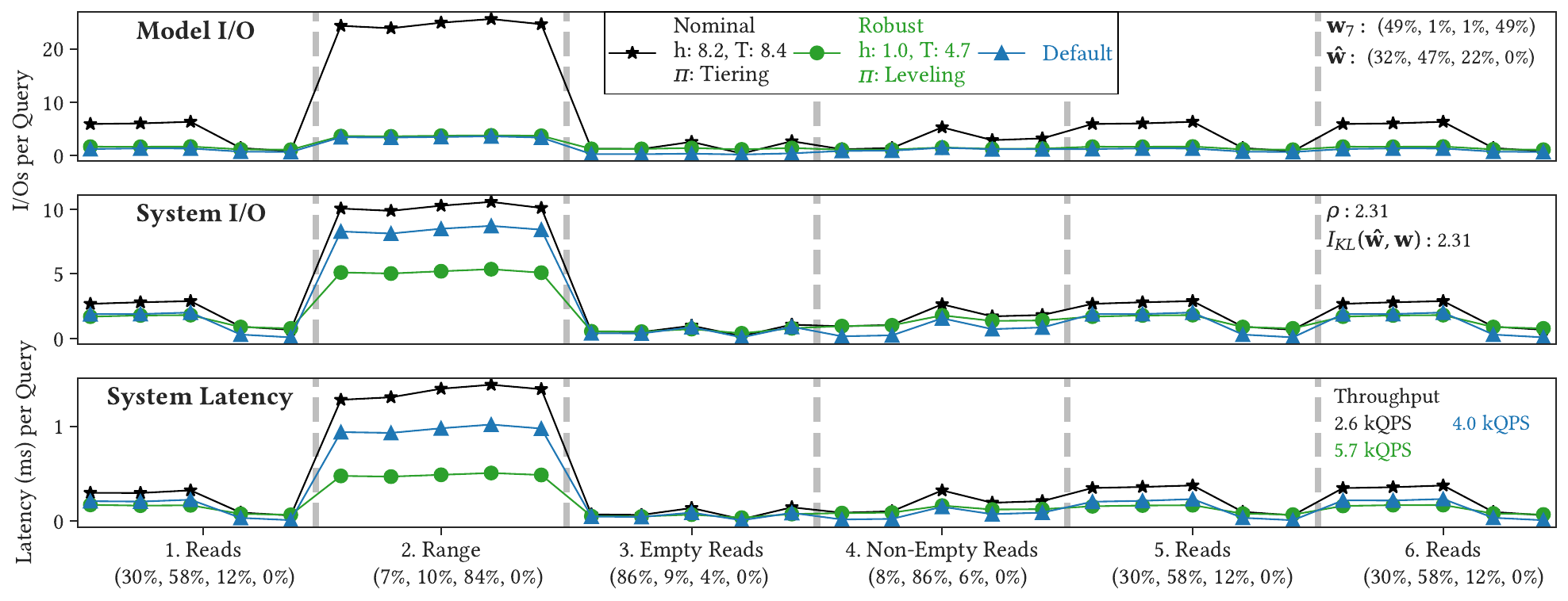}
    \caption{
        System and model performance for robust and nominal tunings in a read-only observed query sequence.
        Both tunings expected a mixed read-write workload.
        The tuning parameter $\rho$ (input uncertainty) matches the observed value of $I_{KL}(\obsworkload, \workload_{7})$ (observed uncertainty).
        Each session contains the label and average workload.
    }
    \label{fig:query_seq_reads_1}
\end{figure*}

\begin{figure*}[t]
    \centering
    \includegraphics[width=\linewidth]{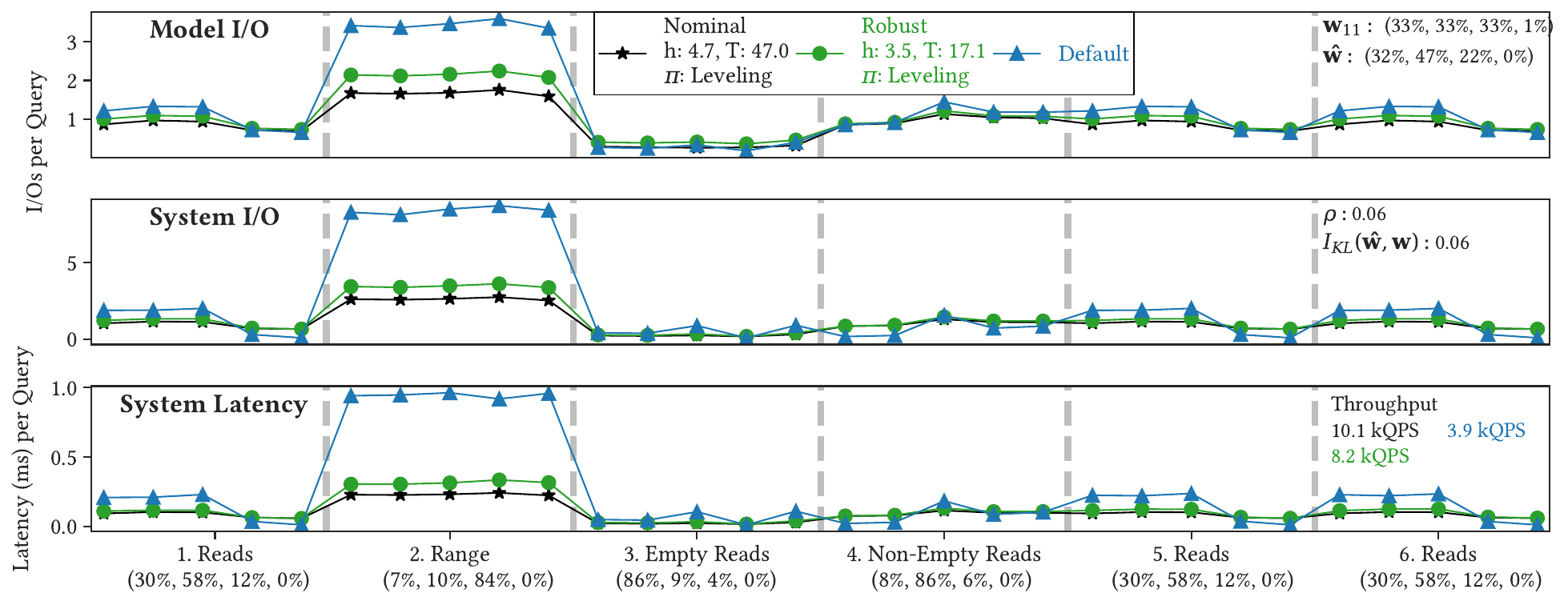}
    \caption{
        Read-only sequence where the observed workloads {\obsworkload} is close to the expected, hence $\rho$ and $I_{KL}(\obsworkload, \workload_{11})$ deviate.
        Both tunings expected a read-heavy workload.
    }
    \label{fig:query_seq_reads_2}
\end{figure*}

\subsection{Experiment Design}

To evaluate the performance of our proposed robust tuning approach, we create multiple instances of RocksDB using different tunings and empirically measure their performance by executing workloads from the uncertainty benchmark {\benchmark}.
To obtain consistent performance metrics, each instantiation of the database is initialized with the same 10 million unique 1 KB key-value pairs.
Each key-value entry has a 16-bit uniformly at random sampled key, with the remaining bits being allocated to a randomly generated value.

While evaluating the performance of the database, we sample a sequence of workloads from the benchmark set {\benchmark}.
Every sampled workload is executed throughout 200,000 queries to measure steady-state performance.
This observation period is sufficient to capture spikes in performance and background compactions allowing us to record accurate performance numbers.
We group sets of workloads into sequences and catalog them into one of six categories — \emph{expected}, \emph{empty read}, \emph{non-empty read}, \emph{read}, \emph{range}, and \emph{write} — based on the dominant query type.
The \emph{expected} session contains workloads with a KL-divergence less than 0.2 w.r.t. the expected workload used for tuning.
For all other sessions, the dominant query type encompasses at least 80\% of the total queries while the remaining queries may belong to any of the remaining types.
When generating keys of the queries to run on the database, we ensure that non-empty point reads will query a key that exists, while empty point reads will query a key that is not present in the database but is sampled from the same domain.
All range queries are generated with minimal selectivity $S_{RQ}$ to act as short-range queries, which on average read zero to two pages per level.
Write queries consist of randomly generated keys that are distinct from the existing keys in the database.
Similarly to Section~\ref{sec:model-evaluation}, we present representative findings for an expected mixed read-write workload ($\workload_{7}$) and an expected read-heavy workload ($\workload_{11}$).

\begin{figure*}[t]
    \centering
    \includegraphics[width=\linewidth]{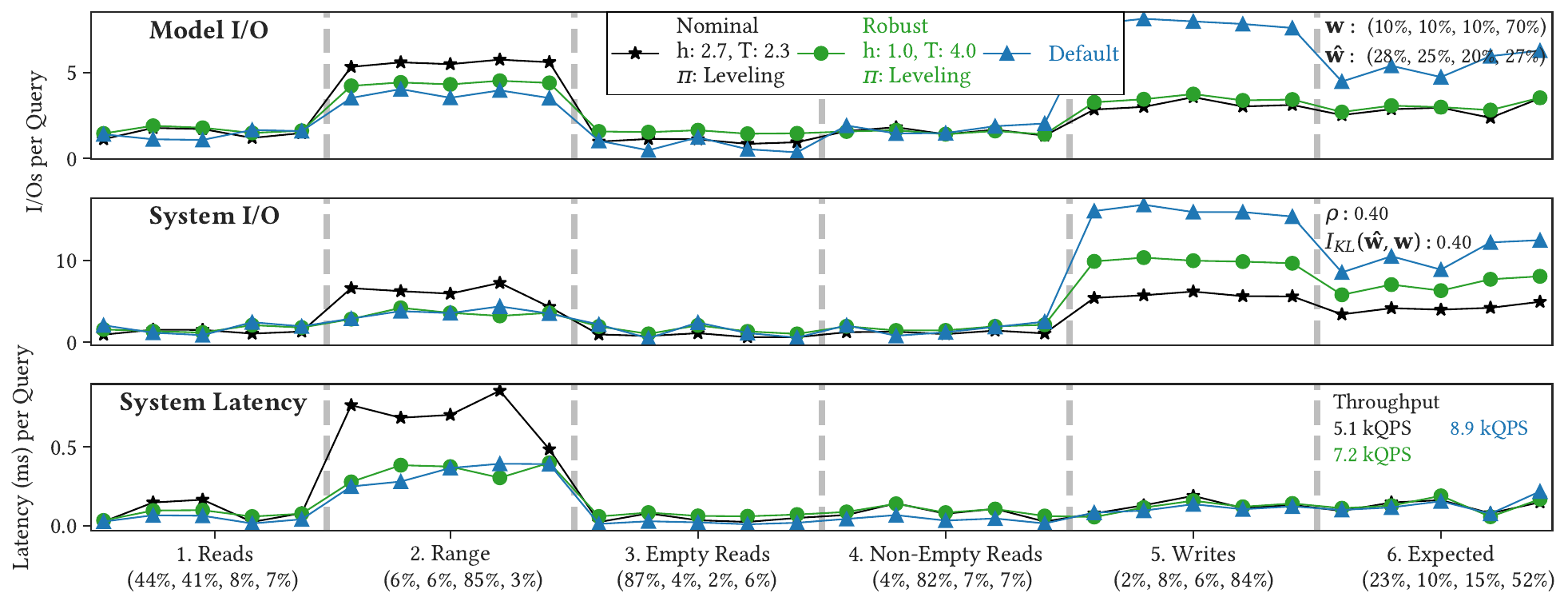}
    \caption{
        Sequence where $\rho$ and $I_{KL}(\obsworkload, \workload)$ closely match.
        Both tunings expected a write-heavy workload.
        Performance fluctuates with writes as it modifies the tree.
    }
    \label{fig:query_seq_hybrid_1}
\end{figure*}

\begin{figure*}[t]
    \centering
    \includegraphics[width=\linewidth]{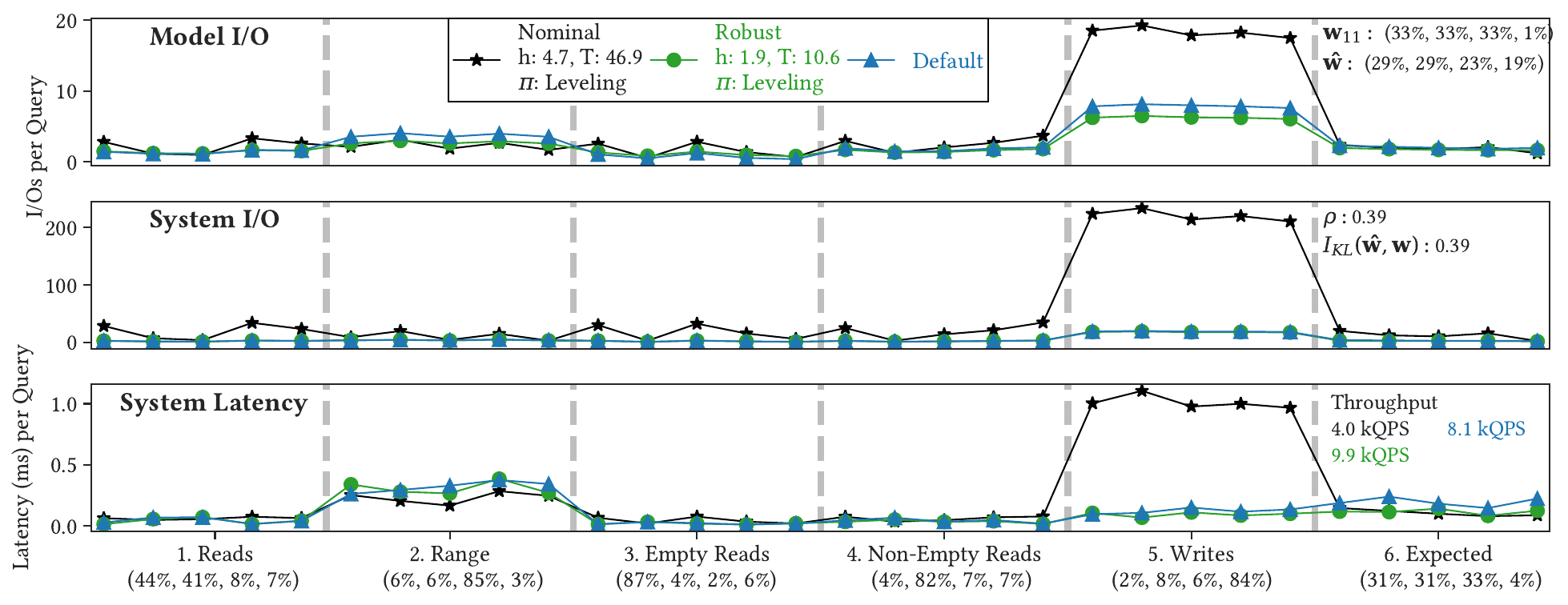}
    \caption{
        Sequence when $\rho$ and $I_{KL}(\obsworkload, \workload_{11})$ closely match.
        Both tunings expected a read-heavy workload.
        System I/O and latency show reductions of up to $90\%$.
    }
    \label{fig:query_seq_hybrid_2}
\end{figure*}

\subsection{Experimental Results}
\label{sec:system-results}

In this section, we replicate key insights from Section~\ref{sec:model-evaluation}, evaluate system performance, and show that {\Endure} scales with database size.
We present detailed results for expected workloads $\workload_{7}$ and $\workload_{11}$.
Table~\ref{tab:throughput_performance} summarizes the normalized delta throughputs $\Delta_\workload(\configuration_N, \configuration_R)$ for all expected workload from $\benchmark$.

\Paragraph{Cost of Tuning}
We first solve for either the nominal or the robust tuning for every experiment.
We note that solving either tuning problem takes less than 10ms, which is negligible w.r.t. workload execution time. 

\Paragraph{Read Performance}
We begin by examining the system performance and verifying that the model-predicted I/O and the system-measured I/O match when considering read queries in Figures~\ref{fig:query_seq_reads_1} and~\ref{fig:query_seq_reads_2}.
In both figures, we include the model-predicted I/Os per query (top), I/Os per query measured on the system (middle), and the system latency (bottom) for nominal, robust, and default configurations across different read sessions.
Additionally, the total throughput numbers in queries per second are reported at the end of the system latency graph.
The empirical measurements confirm that the predicted performance benefits from the model translate to similar performance benefits in practice.
Note that the discrepancy observed between the relative performance between the nominal and the robust tunings in the presence of range queries (session 2 in Figure~\ref{fig:query_seq_reads_1}) is due to the fence pointers in RocksDB.
The analytical model does not account for fence pointers allowing the system to completely skip a run, which may reduce the measured I/Os for short-range queries compared to the predicted I/Os.
Lastly, we consider the default configuration as another tuning of RocksDB that does not take into account workload information.
Therefore, in certain cases such as Figure~\ref{fig:query_seq_reads_1}, it may outperform other configurations. 
This can be explained by the fact that the default configuration includes a larger reserve of memory for the Bloom filter, thereby allowing it to outperform the configurations that expect writes in the executed workload.
However, in other cases such as Figure~\ref{fig:query_seq_reads_2}, we see a large performance dropoff as both the nominal and robust configurations expect a high amount of reads and therefore tune their size ratios accordingly.

\begin{figure}[t]
    \centering
    \includegraphics[width=0.9\linewidth]{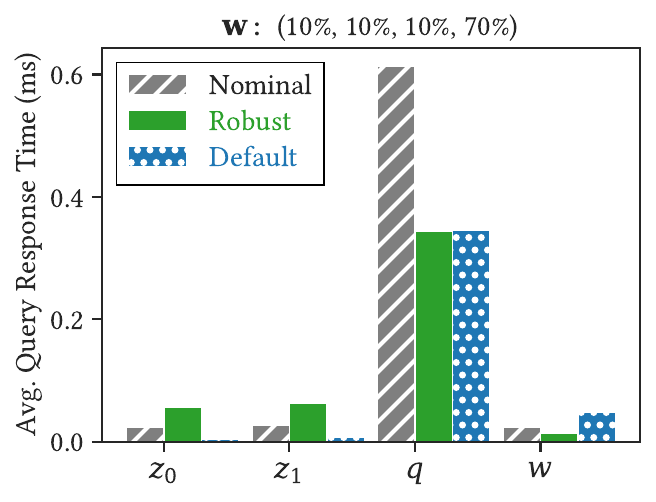}
    \caption{
        Breakdown of the query response time of each operation type for Figure~\ref{fig:query_seq_hybrid_1}.
    }
    \label{fig:query_times_default_w15}
\end{figure}

\Paragraph{Write Performance}
In the presence of writes in Figure~\ref{fig:query_seq_hybrid_2}, the model is still predicting the disk accesses successfully and {\Endure} leads to significant performance benefits.
Note that now the structure of the LSM tree is continually changing across all sessions due to the compactions caused by write queries.
For example, in Figure~\ref{fig:query_seq_hybrid_2} the dip in measured I/Os and latency in the range-query session are the result of empty levels being created via compactions triggered from preceding workloads.
Additionally, writes may appear instantaneous w.r.t. system latency as seen in Figure~\ref{fig:query_seq_hybrid_1} due to RocksDB assigning compactions to background threads.
We observe that the default configuration starts to degrade in performance significantly as more writes are issued to the database.
Figure~\ref{fig:query_times_default_w15} shows the breakdown across each operation type.
As the database experiences more writes, the performance for the default configuration drops off, while both the nominal tuning and robust configurations expect writes and experience a performance improvement.
From the write session of Figure~\ref{fig:query_seq_hybrid_2}, we observe that the nominal tuning suffers from high latency and I/O cost.
This is due to the large size ratio {\sizeratio} that creates a shallow tree with huge levels, causing long stalls during compactions.
Compare this to the robust tuning: the smaller size ratio creates a tree with more stable performance for both I/Os per query and query latency, leading to a higher overall throughput.
Overall, we observe that the robust tuning reduces I/O and latency by up to $90\%$.
Figures~\ref{fig:query_seq_reads_1}--\ref{fig:query_seq_hybrid_2} confirm that our analytical model accurately captures the relative performance of different tunings.

% Table without throughput column
\begin{table}[t]
    \centering\small
    \caption{
        The system measured normalized delta throughputs $\Delta_\workload(\configuration_N, \configuration_N)$ and their respective tunings for experiments on all expected workloads in
        {\benchmark} with an optimally selected $\rho$.
    }
    \begin{tabular}{c ll c}
    \toprule
    Expected & \multicolumn{2}{c}{{\configuration} = ({\sizeratio}, {\mfilt}, {\policy})} & \\
    \cmidrule{2-4}
    Workload ({\workload}) & \multicolumn{1}{c}{${\configuration}_N$} & \multicolumn{1}{c}{${\configuration}_R$} & $\Delta_{\workload}(\configuration_N, \configuration_R)$ \\
    \toprule
    ${\workload_0}$    & $(5.2,\ 3.5,\ \mathbf{L})$ & $(5.1,\ 3.1,\ \mathbf{L})$ & 0.0 \\
    ${\workload_1}$    & $(5.7,\ 9.4,\ \mathbf{L})$ & $(5.0,\ 4.2,\ \mathbf{L})$ & 0.0 \\
    ${\workload_2}$    & $(5.8,\ 5.3,\ \mathbf{L})$ & $(5.0,\ 1.0,\ \mathbf{L})$ & 0.1 \\
    ${\workload_3}$    & $(100,\ 0.0,\ \mathbf{L})$ & $(5.4,\ 1.0,\ \mathbf{L})$ & 0.4 \\
    ${\workload_4}$    & $(17 ,\ 3.2,\ \mathbf{T})$ & $(4.6,\ 1.0,\ \mathbf{L})$ & 1.5 \\
    ${\workload_5}$    & $(5.5,\ 8.8,\ \mathbf{L})$ & $(5.1,\ 3.9,\ \mathbf{L})$ & 0.1 \\
    ${\workload_6}$    & $(63 ,\ 4.8,\ \mathbf{L})$ & $(8.2,\ 1.0,\ \mathbf{L})$ & 0.8 \\
    ${\workload_7}$    & $(8.4,\ 8.2,\ \mathbf{T})$ & $(3.4,\ 1.0,\ \mathbf{L})$ & 0.5 \\
    ${\workload_8}$    & $(62 ,\ 0.0,\ \mathbf{L})$ & $(8.0,\ 1.0,\ \mathbf{L})$ & 0.6 \\
    ${\workload_9}$    & $(8.3,\ 6.9,\ \mathbf{T})$ & $(3.3,\ 1.0,\ \mathbf{L})$ & 0.8 \\
    ${\workload_{10}}$ & $(5.0,\ 0.0,\ \mathbf{L})$ & $(5.0,\ 1.0,\ \mathbf{L})$ & 0.0 \\
    ${\workload_{11}}$ & $(47 ,\ 4.7,\ \mathbf{L})$ & $(11, \ 1.9,\ \mathbf{L})$ & 1.4 \\
    ${\workload_{12}}$ & $(6.2,\ 8.1,\ \mathbf{T})$ & $(2.8,\ 3.1,\ \mathbf{L})$ & 0.2 \\
    ${\workload_{13}}$ & $(5.1,\ 3.5,\ \mathbf{L})$ & $(5.0,\ 1.0,\ \mathbf{L})$ & -0.1\\
    ${\workload_{14}}$ & $(5.1,\ 0.0,\ \mathbf{L})$ & $(5.0,\ 1.0,\ \mathbf{L})$ & -0.1\\
    \bottomrule
    \end{tabular}
    \label{tab:throughput_performance}
    % \vspace{-0.1in}
\end{table}

\Paragraph{Robust Outperforms Nominal for Properly Selected $\rho$}
In the model evaluation (Figure~\ref{fig:rho_vs_rho_hat}), we showed that robust tuning outperforms the nominal tuning in the presence of uncertainty for tuning parameter $\rho$ approximately greater than 0.2.
This is further supported by all the system experiments described.
Specifically, Figures~\ref{fig:query_seq_reads_1} and~\ref{fig:query_seq_hybrid_2} show instances where the KL-divergence of the observed workload averaged across all the sessions w.r.t. the expected workload is close to the tuning parameter $\rho$.
Additionally, we present results for all expected workloads in Table~\ref{tab:throughput_performance}.
Each entry in the table summarizes the total throughput after running the same experimental setup presented in Figure~\ref{fig:query_seq_hybrid_2}.
We observe that the robust outperforms the nominal in 10 of our expected workloads, with only 2 workloads where robust tuning does worse, however, in these cases the reported throughputs are comparable.
In each of these experiments, the robust tuning outperforms the nominal resulting in up to a $90\%$ reduction in latency and system I/O.
Lastly, in Figure~\ref{fig:query_seq_reads_2}, the observed workloads are similar to the expected one ($I_{KL}(\obsworkload, \workload_{11}) < 0.2$), resulting in a latency increase of $20\%$.

\begin{figure}[t]
    \centering
    \includegraphics[width=0.9\linewidth]{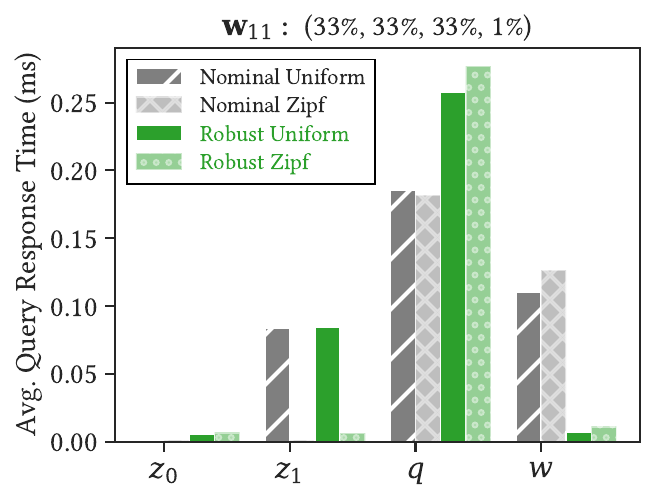}
    \caption{
        Query times for each operation (empty reads, non-empty reads, range reads, and writes) with an expected workload $\workload_{11}$.
        Robust tunings were generated with $\rho = 1$.
    }
    \label{fig:query_times_w11}
\end{figure}

\Paragraph{Balancing Query Times}
To determine how the tunings from {\Endure} outperform the nominal tunings we analyze the query response times for each operation for an expected workload $\workload$.
We observe that robust tunings will provide lower performance for the queries that dominate the expected workload, however, as a tradeoff these tunings perform exceptionally well in unexpected operations.
For example, Figure~\ref{fig:query_times_w11} shows robust tuning performs worse in both range queries and empty point queries, however, in exchange we observe a significant decrease in the response time of write queries.
Hence, if the executed workload contains writes, robust tuning will increase overall throughput, especially in the presence of write spikes.

\Paragraph{Workload Skew}
To verify that {\Endure} works across workload distributions, we break down the different query response times in Figure~\ref{fig:query_times_w11}.
When the keys generated for the workload follow a Zipfian distribution, we see that the response time for non-empty read queries is significantly lowered.
This is in part due to keys towards the top of the tree being repeated, therefore the query does not need to traverse further down the tree resulting in an increase in false positives from the Bloom filter.
Regardless, we observe the same patterns for uniform and Zipfian distributions; {\Endure} tunings achieve a better tradeoff in performance for the dominant query types in the expected workload with that of other query types, thereby preventing large performance regressions.

\begin{figure}[t]
    \centering
    \includegraphics[width=\linewidth]{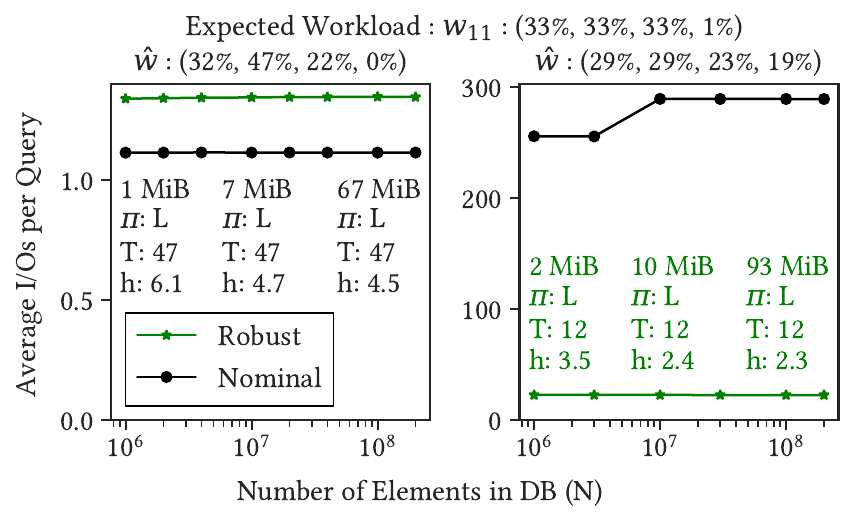}
    \caption{
        Impact of database size on performance.
        All tunings use the same expected workload $\workload_{11}$ with executed workloads shown above each graph.
        Points at each power of 10 show $m_{buf}$ and the tuning $\configuration$ (L for leveling, T for tiering).
    }
    \label{fig:perf_db_size}
\end{figure}

\Paragraph{ENDURE Scales with Data Size}
To verify that {\Endure} scales, we repeat the previous experiments, while varying the size of the initial database.
Each point in Figure~\ref{fig:perf_db_size} is calculated based on a series of workload sessions similar to the ones presented in Figures~\ref{fig:query_seq_reads_2} (\ref{fig:query_seq_hybrid_2}) for the left (right) part of Figure~\ref{fig:perf_db_size}.
All points use the same expected workload, therefore the nominal and robust tunings are the same across each graph.
We observe that the robust and nominal tuning increases buffer memory as the initial database size grows.
As a result, for all cases, the number of initial levels is the same regardless of the number of entries.
This highlights the importance of the number of levels w.r.t performance.
Finally, the performance gap between robust and nominal stays consistent as database size grows, showing that {\Endure} is effective regardless of data size.

% =============================================================================
\section{Robustness of Flexible Designs}
\label{sec:robustness-flexible-models}
% =============================================================================
In this section, we explore the nuanced differences between \emph{flexibility} and \emph{robustness} of LSM designs.
Under an ideal scenario, where the expected workload is known a priori to tuning, the flexibility of K-LSM provides a high-performance benefit, however, this benefit vanishes in scenarios where the executed workload changes from the expected.
This observation is in line with the intuition that {\Endure}'s robust tunings proactively compensate for potential changes in the workload distribution post-deployment, thereby providing better performance in situations where the executed workload differs from the expected.

\subsection{Experiment Design}
To evaluate the robustness of all designs, we design an experiment to compare different optimal tunings for a subset of designs listed in Table~\ref{tab:lsm_variations}.
For each expected workload in the uncertainty benchmark in Table~\ref{tab:expected-workloads}, we obtain a list of tunings by solving {\nominal} for various LSM data layouts.
We also compute a robust tuning with input $\robustrho = 2$.
Next, for every observed workload in {$\benchmark$}, we calculate the KL-divergence w.r.t. the expected workload used to obtain each tuning and plot the average I/Os per query for this observed workload ($\cost(\obsworkload, \configuration)$) versus the KL-divergence.
We then execute the initial expected workload for all designs, reset the database to the initial state, and then progressively repeat this process for workloads further away from the expected workload.
Similar to Section~\ref{subsec:nominal-experiment}, we adopt the following setting for system parameters: the database initially holds 10 billion with each entry at 1 KB; page size is 4 KB; and memory budget is set to 10 bits per element or a total of 10 GB.

\subsection{Comparison Results}

\begin{figure}
    \centering
    \includegraphics[width=\columnwidth]{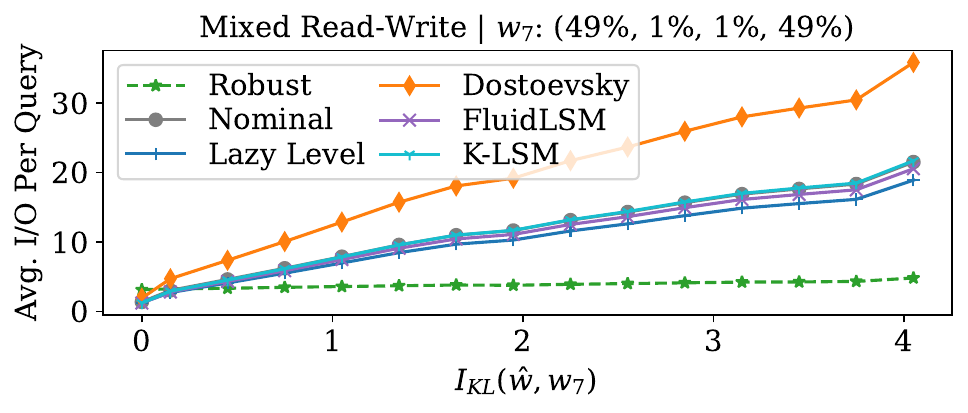}
    \includegraphics[width=\columnwidth]{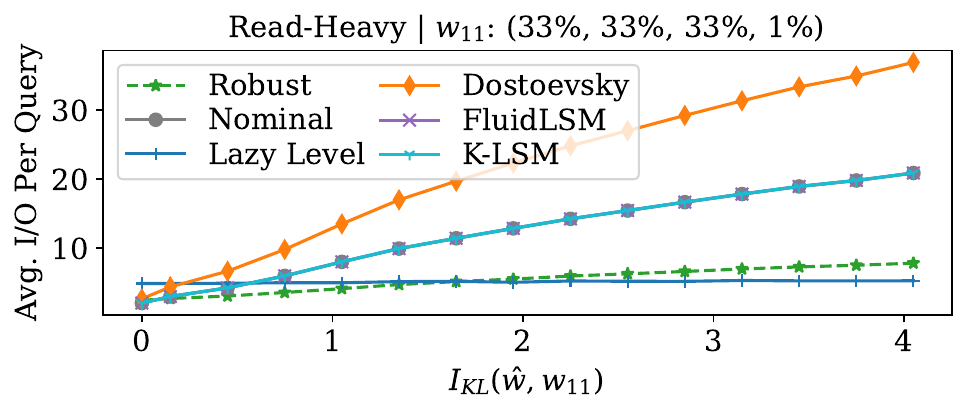}
    \caption{
        The cost of each LSM model as the observed workload {\obsworkload} drifts away from the expected workload {\workload}.
    }
    \label{fig:distance-vs-cost}
\end{figure}

Figure~\ref{fig:distance-vs-cost} shows the average I/O per query for various LSM models.
Note that for $\workload_{11}$, the performance lines for {\kmodel}, {\fluidmodel}, and {\Endure}'s nominal tuning slightly overlap, as the configurations are the same.
The same occurs for {\kmodel} and {\fluidmodel} on $\workload_{7}$.
When evaluating Dostoevsky, we fix memory allocation such that the buffer size is kept at 2 MB as per convention, while the remaining memory is delegated to storing Bloom filters~\cite{Dayan2018,RocksDB2021}.

\Paragraph{Robustness of Various LSM Designs}
It should be noted that in instances where the observed workload closely matches the expected workload, {\Endure}'s robust tunings underperform. 
This is consistent with previous experiments and the intuition that robust tunings proactively compensate for potential changes in the observed workload distribution.
As the observed workload drifts further from the expected, robust tunings maintain consistent performance while other tunings show a steady increase in the average number of I/Os.
This observation can be attributed to the selected tuning.
For example, with $\workload_{7}$ models such as {\dostmodel} ($T = 47$, all $K_i = 1 $) and {\kmodel} ($\mfilt = 4.4$ bits per entry, $T = 48$, all $K_i = 1$) optimally selects larger size ratios with effectively leveling policies to accommodate for the expected high amount of writes.
In contrast, the robust tuning ($T = 9$, $\policy = L$) selects a size ratio that performs reasonably well in comparison, however, the selected size ratio is small enough to accommodate a large shift to reads.

In the presence of workload drifts, we observe that most models, except {\llmodel}, experience a performance degradation similar to {\Endure}'s nominal tuning.
The optimal tuning of {\llmodel} ($\mfilt = 4.6$ bits per entry, $T = 2$) performs robustly when tuned for a read-heavy workload ($\workload_{11}$).
It should be noted that the optimizer selects a size ratio $T = 2$ that enables {\llmodel} to accommodate an increase in writes, as merge operations are relatively cheap. 
Furthermore, increasing the size ratio any further could lead to the creation of upper levels that follow a tiering policy, thereby degrading the performance.

\section{Discussion}
In this section, we discuss the key insights gained from benchmarking and
testing {\Endure}.
\newline

\Paragraph{Robustness is All You Need}
When deploying LSM trees, it is evident that tuning with some knowledge about the workload can improve performance, but accounting for uncertainty in the tuning process can provide an even greater benefit for performance in the long run.
To support this, in Section~\ref{sec:system-results}, we show that the cost model can accurately predict the empirical measurements.
Then using our model, we compared over 700 different robust tunings with their nominal counterparts over the uncertainty benchmark set {\benchmark}, leading to approximately 8.6 million comparisons.
Robust tunings comprehensively outperform the nominal tunings in over 80\% of these comparisons.
We further cross-validated the relative performance of the nominal and the robust tunings in over 300 comparisons using RocksDB.
The empirical measurements overwhelmingly confirmed the validity of our analytical models, and the few instances of discrepancy in the scale of measured I/Os, such as the ones discussed in previous sections, are easily explained based on the structure of the LSM tree.

\Paragraph{Leveling is ``more'' Robust than Tiering}
One of the key takeaways of applying robust tuning to LSM trees is that \emph{leveling is inherently more robust} to perturbations in workloads when compared to pure tiering.
Note that this is evident from Table~\ref{tab:throughput_performance}, where all robust tunings suggest leveling as the compaction policy.
This observation is in line with the industry practice of deploying leveling or hybrid leveling over pure tiering.

\Paragraph{Robustness is Not Inherent}
As evident in Figure~\ref{fig:distance-vs-cost}, the final takeaway when evaluating robust tunings compared to optimal tuning of other flexible models is that \emph{robustness is not inherent to a model and must be considered in the tuning process}.
We observe that flexible models may provide better initial performance, however, only {\Endure}, which explicitly accounts for workload uncertainty in the tuning process, performs well w.r.t. to a changing workload.
While other models may exhibit some degree of robust performance in specific and limited scenarios, only the robust tuning consistently performs well in the presence of workload drift across all different expected workloads.
Based on our analytical and empirical results, we recommend that robust tuning should always be employed when tuning an LSM tree unless the future workload distribution is known with an absolute certainty.

\Paragraph{Limitations}
One of the key challenges during the evaluation of tuning configurations in the presence of uncertainty is in measuring steady-state performance.
Background compactions create variability in performance requiring longer database testing sessions to see accurate performance numbers. 
To observe trends across multiple tunings we had to strike a balance between exhaustive testing and runtime.
Using off-the-shelf optimizers, such as the SLSQP solver from SciPy mentioned in Section~\ref{sec:tuning-lsm-trees}, present restrictions in terms of the complexity of designs that we can optimally tune.
Numerical solvers are sensitive to hyperparameters such as starting conditions and step size.
Therefore, tuning performance can greatly vary based on the correct initialization of such hyperparameters.
Furthermore, the stability and accuracy of numerical solvers suffer with an increase in the number of decision variables.
We observed that when solving the {\robustw} problem for the most flexible designs, the combination of hyperparameter sensitivity and numerical instability with additional decision variables leads to suboptimal solutions.

While we have deployed and tested robust tuning on LSM trees, the robust paradigm of {\Endure} is a generalization of a minimization problem that is at the heart of any database tuning problem.
Hence, similar robust optimization approaches can be applied to \emph{any database tuning} problem assuming that the underlying cost model is known, and each cost model component is convex or can be accurately approximated by a surrogate.

% =============================================================================
\section{Related Work}
\label{sec:related_work}
% =============================================================================
Database tuning is a notoriously hard problem, however, there has been a plethora of recent research.
We provide a discussion of related works around the field of data systems tuning.
\newline

\Paragraph{Tuning Data Systems}
Database systems are notorious for having numerous tuning knobs. 
These tuning knobs control fine-grained decisions (e.g., number of threads, amount of memory for bufferpool, storage size for logging) as well as basic architectural and physical design decisions about partitioning, index design, materialized views that affect storage and access patterns, and query execution~\cite{Bruno2005,Chaudhuri1998a}.
The database research community has developed several tools to deal with such tuning problems.
These tools can be broadly classified as offline workload analysis for index and views design~\cite{Agrawal2004,Agrawal2000,Chaudhuri1997,Dageville2004,Valentin2000,Zilio2004}, and periodic online workload analysis~\cite{Bruno2006,Schnaitter2006,Schnaitter2007,Schnaitter2012} to capture workload drift~\cite{Holze2010}.
In addition, there has been research on reducing the magnitude of the search space of tuning~\cite{Bruno2005,Dash2011} and on deciding the optional data partitioning~\cite{Athanassoulis2019,Papadomanolakis2004,Serafini2016,Sun2014,Sun2016}.
These approaches assume that the input information about resources and workload is accurate. 
When it is proved to be inaccurate, performance is typically severely impacted.

\Paragraph{Adaptive \& Self-designing Data Systems}
A first attempt to address this problem was the design of \textit{adaptive} systems which had to pay additional transition costs (e.g., when deploying a new tuning) to accommodate shifting workloads~\cite{Idreos2007,Graefe2010a,Graefe2010c,Schuhknecht2018}. 
More recently the research community has focused on using machine learning to learn the interplay of tuning knobs, and especially of the knobs that are hard to analytically model to perform cost-based optimization. 
This recent work on self-driving database systems~\cite{Aken2017,Ma2018,Pavlo2017} or self-designing database systems~\cite{Idreos2019,Idreos2019a,Idreos2018a,Idreos2018} is exploiting new advancements in machine learning to tune database systems and reduce the need for human intervention, however, they also yield suboptimal results when the workload and resource availability information is inaccurate.

\Paragraph{Robust Database Physical Design}
One of the key database tuning decisions is physical design, that is, the decision of which set of auxiliary structures should be used to allow for the fastest execution of future queries. 
Most of the existing systems use past workload information as a representative sample for future workloads, which often leads to suboptimal decisions when there is significant workload drift. 
Cliffguard~\cite{Mozafari2015} is the first attempt to use unconstrained robust optimization to find a robust physical design. 
Their method is derived from Bertsimas et al. in~\cite{Bertsimas2010-a}, a numerical optimization approach using alternating gradient ascent-descent to optimize problems without closed-form objectives.
In contrast to Cliffguard, {\Endure} focuses on the LSM tree tuning problem which uses an analytical closed form objective in Equation~\eqref{eq:thecost}.
This allows us to directly solve a Lagrangian dual problem instead of relying upon numerical optimization techniques.
Furthermore, we found that the approach in Cliffguard, when applied to our objective, fails to converge even after an extensive hyperparameter search.

% =============================================================================
\section{Conclusion}
\label{sec:conclusion}
% =============================================================================

In this work, we explored the impact of workload uncertainty and LSM design flexibility on the performance of LSM tree databases.
Based on our explorations, we introduce {\Endure} -- a robust tuning paradigm that recommends robust designs to mitigate performance degradation under scenarios of deviating workloads.
We showed that in the presence of uncertainty, {\Endure} increases database throughput compared to standard tunings by up to $5 \times$.
Furthermore, we proposed a unified LSM design with an associated flexible cost model that can express multiple LSM data layout designs, and provide evidence that our cost model closely matches the behavior measured on a database system.
We used this cost model to analyze the robustness of flexible models and provide evidence that \emph{robustness is not inherent} to a particular design, rather it must be an important consideration during the tuning process.
Through both model-based and extensive experimental evaluation of {\Endure} within the state-of-the-art LSM-based storage engine RocksDB, we show that the robust tuning methodology consistently outperforms classical tuning strategies.
{\Endure} can be an indispensable tool for database administrators to evaluate deployed tunings performance, as well as recommend optimal tunings without resorting to expensive database experiments.

\section*{Acknowledgements}
We thank Anwesha Saha and Sakshi Sharma for their help in the experimental analysis and the anonymous reviewers for their valuable feedback.
This work is partially funded by an IBM Ph.D. Fellowship, a Red Hat Incubation Award, a Meta gift, and NSF Grants No. IIS-1813406, No. IIS-1908510, and No. IIS-2144547.

\balance
\bibliographystyle{spmpsci}
\bibliography{library-no-url,bibliography}

\end{document}